\documentclass[conference]{IEEEtran}

\usepackage{times}
\usepackage{epsfig}
\usepackage{graphicx}
\usepackage{amsmath}
\usepackage{amssymb}
\usepackage{tabulary}
\usepackage{multirow}
\usepackage{morefloats}
\usepackage{subfigure}
\usepackage{floatrow}
\usepackage{array}
\usepackage{mathtools}
\usepackage{textcomp}

\usepackage[ruled,norelsize]{algorithm2e}

\makeatletter
\newcommand{\removelatexerror}{\let\@latex@error\@gobble}
\makeatother

\makeatletter
\let\ftype@table\ftype@figure
\makeatother

\usepackage[pagebackref=true,breaklinks=true,letterpaper=true,colorlinks,bookmarks=false]{hyperref}

\begin{document}

\title{Creating Simplified 3D Models with High Quality Textures}

\author{Song Liu, Wanqing Li, Philip Ogunbona, Yang-Wai Chow\\
Advanced Multimedia Research Lab \\
University of Wollongong, Wollongong, NSW, Australia, 2522\\
{\tt\small \{sl796,wanqing,philipo,caseyc\}@uow.edu.au}
}

\maketitle

\begin{abstract}
 This paper presents an extension to the KinectFusion algorithm which allows
creating simplified 3D models with high quality RGB textures.
 This is achieved through
 (i) creating model textures using images from an HD RGB camera that is
calibrated with Kinect depth camera,
 (ii) using a modified scheme to update model textures in an asymmetrical
colour volume that contains a higher number of voxels than that of the geometry
volume,
 (iii) simplifying dense polygon mesh model using quadric-based mesh decimation
 algorithm, and
 (iv) creating and mapping 2D textures to every polygon in the output 3D model.
 The proposed method is implemented in real-time by means of GPU parallel
processing. Visualization via ray casting of both geometry and colour volumes
provides users with a real-time feedback of the currently scanned 3D model.
 Experimental results show that the proposed method is capable of keeping
the model texture quality even for a heavily decimated model and that, when
reconstructing small objects, photorealistic RGB textures can still be
reconstructed.
\end{abstract}

\section{Introduction}

Generating 3D models based on real-world environments and with high quality
textures is of great significance to many fields including civil
engineering, 3D printing, game design, movie, virtual reality and preservation
of cultural heritage artefacts.
Various computer vision-based approaches have been proposed to create 3D models
and deal with the associated classical problems such as simultaneous
localization and mapping (SLAM), and structure-from-motion (SFM).
To date, impressive progress has been made in this domain
\cite{Chiuso003-dmotion}\cite{Agarwal:2011:BRD:2001269.2001293}\cite{
conf/icra/EndresHESCB12}. Largely, many approaches use visual key
points to build 3D models leading to sparse point cloud based 3D reconstruction.
Conventional dense reconstruction method \cite{990921}\cite{5580584} on the
other hand usually require professional sensors such as high-fidelity laser
scanners or time-of-flight (ToF) depth cameras which are very expensive.

The release of commodity RGB-D cameras such as the
{Microsoft Kinect\texttrademark} and {Asus
Xtion\texttrademark} has made dense 3D reconstruction possible at an affordable
cost. This, along with the KinectFusion algorithm
\cite{Newcombe:2011:KRD:2120094.2120179}\cite{Izadi:2011:KRR:2047196.2047270} ,
 has enabled real-time dense 3D reconstruction using a low-cost RGB-D
camera and GPU parallel processing.
Subsequent efforts by other researchers have led to the development of several
KinectFusion-based methods
\cite{Whelan12rssw}\cite{6751517}\cite{Bylow-RSS-13}\cite{
NieBner:2013:RRS:2508363.2508374} that allow efficient 3D reconstruction on a
large scale and with higher reconstruction quality.
However, current KinectFusion-based methods tend to deliver 3D models with high
quality geometry but low quality texture. In other words, the works on improving
model textures are less advanced.
Moreover, 3D models created by dense 3D reconstruction usually contain
significant redundant information,
which would increase the model complexity and lower the rendering efficiency.
In many cases, it is necessary to simplify a dense 3D model to achieve higher
rendering efficiency, especially for large scale rendering or on platforms with
limited processing powers such as cell phones and tablets.
It is noteworthy that for many 3D models generated from existing 3D
reconstruction systems, the quality of model texture is directly related to the
model complexity. Furthermore, decimation of the polygon mesh models will
degrade the model texture.

\section{Related Work}

The problem of reconstructing geometry and texture of real world
has remained an active challenge in the field of computer vision for decades.
We now review some of the extant aproaches and the associated results.

Conventional 3D reconstruction approaches usually do not
consider model texture information, or represent model texture in a simple way.
Chen and Medioni \cite{Chen1992} average overlapping range
images and connect points based on simple surface topology to create polygon
mesh models; model textures are totally ignored.
Turk and Levoy \cite{Turk:1994:ZPM:192161.192241} propose
mesh zippering as an extension to Chen and Medioni's work; they stitch
polygon meshes to create 3D model without textures.
Some point-based 3D reconstruction
methods
\cite{Rusinkiewicz:2002:RMA:566654.566600}\cite{5457479}\cite{
Henry10rgbdmapping:}\cite{6343050} use simple unstructured point representations
that are directly captured from many range imaging devices.
These methods do not model connected surfaces which usually
requires post processing to generate polygons.
In most popular point-based 3D model rendering
techniques
\cite{Gross:2007:PG:1202384}\cite{Pfister:2000:SSE:344779.344936}\cite{
Zwicker:2001:SS:383259.383300}, textures are simply represented by colours
attached to each point in the model.

The release of low-cost RGB-D cameras like the
{Microsoft Kinect\texttrademark} and
{Asus Xtion\texttrademark} opens up new opportunities to 3D reconstruction
in terms of providing easy access to depth imaging.
KinectFusion
\cite{Newcombe:2011:KRD:2120094.2120179}\cite{Izadi:2011:KRR:2047196.2047270}
adopts volumetric data structure to store reconstructed scene surface
\cite{Curless1996}\cite{Hilton1996} and realise real-time reconstructions using
GPU.
Although model textures are not considered in the original
KinectFusion algorithm, it inspires multiple volumetric 3D reconstruction
methods using commodity RGB-D cameras that try to create dense polygon mesh with
RGB textures.
Among these KinectFusion-based methods, an open source
C++ implementation of KinectFusion from Point Cloud Library (PCL)
\cite{PCL:web}, Whelan et al. \cite{Whelan12rssw} and Bylow et al.
\cite{Bylow-RSS-13} use a colour volume to store and update RGB
texture
information. In their methods, model textures on reconstructed 3D models are
represented by colours on model vertices, and these colours are linearly
interpolated within each polygon.
This popular 3D model texture representation can also be easily found
in many other 3D reconstruction methods
\cite{6751517}\cite{Whelan13icra}\cite{NieBner:2013:RRS:2508363.2508374}.
While the texture representation is straightforward and easy to implement,
simplifying the model inevitably degrades the texture quality because it is
determined by the number of vertices in the model.

Zhou and Koltun \cite{Zhou:2014:CMO:2601097.2601134} reconstruct 3D
models with high quality colour textures using a commodity depth camera and an
HD RGB camera. HD model textures are refined using optimized camera poses in
tandem with non-rigid correction functions for all images.
Because model textures are also represented by colours assigned to each vertex,
the generation of high quality textures requires increasing the number of
vertices and polygons of a 3D model. This results in increased model
complexity.

In order to create simplified 3D models, Whelan et al.~\cite{Whelan14ras}
decimate dense polygon mesh model by means of
planar simplification and simultaneously preserve model colour information
by 2D texture mapping. However, their work focuses on scene reconstruction where
flat floors and walls tend to be over represented by millions of polygons.
For 3D models of arbitrary shapes, planar simplification is not suitable.
Moreover, their texture preserving method does not consider generating
HD RGB textures with resolution beyond what a Kinect RGB camera currently
provides.

In this paper, a KinectFusion-based method for creating simplified
3D models with high quality textures is presented.
Texture information is stored and updated in a colour volume with a
higher dimension than that of the truncated signed distance function (TSDF)
volume. Two-dimensional texture images are extracted from colour volume and are
mapped to reconstructed 3D models so that model texture can retain its quality
even on a simplified 3D model with much fewer number of polygons.

\section{Background}

Our method is based on the open source C++ implementation of the KinectFusion
algorithm provided by PCL~\cite{PCL:web}.
In KinectFusion, retrieved RGB image and registered depth map are taken as inputs.
A vertex map and normal map pyramid based on input depth map is
calculated; the value is then used to estimate the camera motion using
iterative closest point (ICP) algorithm \cite{Besl:1992:MRS:132013.132022}
in conjunction with a predicted surface vertex map. The map is derived from
currently stored depth information.
With estimated camera motions, depth and RGB information from the input
frame is used to update the current model.
In the work of KinectFusion, scanned depth information is stored in a
TSDF volume, in which each voxel stores the
distance to its closest surface.
Since TSDF volume contains geometric information of scanned 3D model
and is updated using depth map, in our method, it is referred to as the
geometry volume. The predicted surface vertex map used for camera tracking is
obtained by ray tracing the geometry volume.
The original KincetFusion does not consider reconstructing model textures.
Its C++ implementation in PCL capture and save texture information
using a separate colour volume that has the same dimension and size as
the TSDF volume.
In our paper, this colour integration method is adopted and extended.
Given updated geometry and colour volumes, 3D polygon mesh model with
textures can be extracted using marching cube algorithm
\cite{Lorensen87marchingcubes:}; a method that aims to produce triangle
polygons for visualization.

\section{Improved Method}

\begin{figure*}[h!t]
\centering
\includegraphics[width=5in]{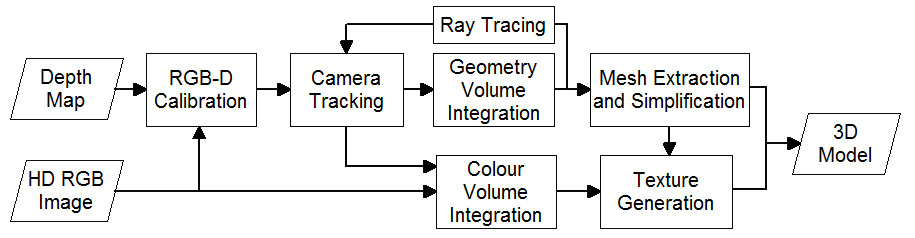}
\caption{Block diagram showing the
workflow of improved method.}
\label{3dm_flow}
\end{figure*}

The workflow of our improved method is shown in Fig~\ref{3dm_flow}. The
similarity with the KinectFusion process is noticeable. However, in order to
create simplified 3D models with high quality textures, the following major
improvements are made:
\begin{itemize}
\item HD RGB camera is added to achieve higher model texture quality;
\item Colour volume integration scheme is revised to
provide an asymmetrical colour volume;
\item Dense polygon mesh model is simplified without losing much
geometry information;
\item Simplified model is textured and details are retained.
\end{itemize}
In subsequent subsections each major improvement is presented in more detail.

\subsection{HD Camera Setup and RGB-D Calibration}

\begin{figure}[ht!]
\centering
\includegraphics[width=2.5in]{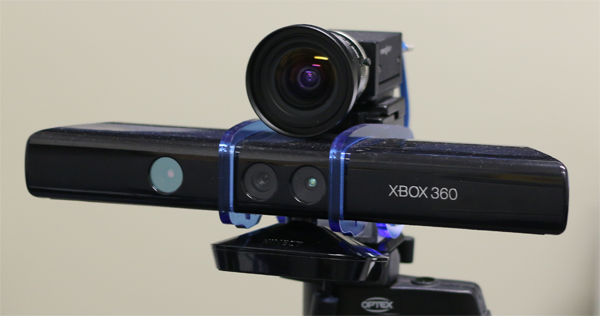}
\caption{Kinect sensor with HD RGB camera}
\label{HD_Kin}
\end{figure}

The on-board RGB camera of the Kinect sensor can only return RGB images
in VGA resolution ($640 \times 480$) and this insufficient
for generating high quality textures. This shortcoming is mitigated,
in order to achieve high quality texture mapping, by rigidly attaching an
external HD RGB camera to the Kinect sensor assembly. The resulting assembly
delivers high-definition RGB images (see Fig \ref{HD_Kin}).

The HD RGB image and depth map are misaligned due to the different
placements and field of views of the HD RGB camera and the depth camera of
Kinect sensor. Hence high quality texture mapping will depend on accurate
mapping relation between HD RGB camera space and Kinect depth camera
space.

Herrera C's RGB-D calibration toolbox \cite{6205765} is adopted to
calculate camera intrinsic and extrinsic parameters. Camera intrinsic
parameters include camera principal point (image centre) and focal length in
pixel-related units which are used to project points between image space and
camera space. Camera intrinsic parameters can be compactly represented in
a matrix, $K \in \mathbb{R}^{3\times3}$.
Let the coordinate vector in image space be denoted by $(x, y, z)^T$ and
the coordinate vector in the camera space be $(X, Y, Z)^T$. The transformation
from camera space to image space can be written as
\begin{equation}
(x, y, z)^T=K(X, Y, Z)^T.
\label{img2cam}
\end{equation}
Camera extrinsic parameters describe the relative position
of the camera to another object (another camera in our case),
and can be written as a rotation plus a translation. The extrinsic parameters
from Kinect depth camera space to HD RGB camera space are denoted
by rotation matrix $R_{calib} \in \mathbb{R}^{3\times 3}$ and translation
vector $t_{calib}\in \mathbb{R}^{3}$.
Let the intrinsic parameters of Kinect depth camera and HD
RGB camera be donoted as $K_c \in \mathbb{R}^{3\times 3}$ and $K_{hd} \in
\mathbb{R}^{3\times 3}$ respectively.

Given calibrated camera intrinsic and extrinsic parameters, a mapping
from depth image to HD RGB image is calculated as follows.
Assume a point in real world is captured by the depth camera whose
coordinate in depth image space is $(x_c, y_c, z_c)$.
Let its 3D coordinate in depth camera space be denoted as $(X_c, Y_c, Z_c)$;
the 3D coordinate in HD RGB camera space as $(X_{hd}, Y_{hd}, Z_{hd})$; and its
coordinate in HD RGB image space as $(x_{hd}, y_{hd}, z_{hd})$.
The equations governing the respective transformations can be written as
follows:
\begin{equation}
(X_c, Y_c, Z_c)^T=K_{c}^{-1}(x_c, y_c, z_c)^T;
\label{ci2cc}
\end{equation}
\begin{equation}
(X_{hd}, Y_{hd}, Z_{hd})^T=R_{calib}(X_c, Y_c, Z_c)^T+t_{calib};
\label{c2hd}
\end{equation}
\begin{equation}
(x_{hd}, y_{hd}, z_{hd})^T=K_{hd}(X_{hd}, Y_{hd}, Z_{hd})^T.
\label{hdc2hdi}
\end{equation}
The mapped 2D coordinate $(x_{hd}, y_{hd})$ in HD RGB image will be considered when updating colour volume.

\subsection{Colour Volume Integration}

In the implementation of KinectFusion from PCL, RGB texture information
is saved in a separate volume but similarly dimensioned and
sized as the TSDF geometry volume. This establishes one-to-one correspondence
between voxels in colour volume and geometry volume.
In our method, modifications are made to improve PCL's technique to deliver higher quality model textures.

Since all texture information is saved in the colour volume,
a colour volume with higher dimension will result in better texture quality.
As mentioned earlier, most KinectFusion-based methods tend to deliver 3D models
with high quality geometry but low quality textures.
Based on this observation and limited GPU memory availability, a strategy to
achieve higher colour volume dimension is to make the
geometry and colour volumes asymmetrical.
Specifically, the dimension of colour volume is made higher than that of the
geometry volume while their sizes remain the same.
In this way, the actual space in real world taken by a colour voxel is much
smaller than that is taken by a geometry voxel; this results in capturing more
textures details.

The colour volume containing updated texture information from the $1^{st}$
frame to the $n^{th}$ frame is denoted by $C_n(v)$, where $v\in\mathbb{N}^3$ is
the 3D coordinate representing the location of a voxel in the colour volume.
Each voxel of the colour volume stores a $3\times1$ RGB colour vector $C_n(v)$ and a weight $W_n(v)$.

When updating the colour volume,
for each voxel $v$ in the colour volume, its 3D coordinates $(X_v, Y_v, Z_v)$
in the depth camera space of the first frame is considered. Given all camera
motions between consecutive frames so far, by synthesizing all the motions, the
camera motion $(R_n, t_n)$ from the first frame to the current $n^{th}$ frame
can be calculated.  Therefore, the coordinates of this voxel in the current
camera coordinate system, $(X_{vc}, Y_{vc}, Z_{vc})$, can be calculated as
\begin{equation}
(X_{vc}, Y_{vc}, Z_{vc})^T=R_n(X_v, Y_v, Z_v)^T+t_n.
\label{v2c}
\end{equation}
Using the depth camera intrinsic matrix $K_c$, 3D point $(X_{vc}, Y_{vc},
Z_{vc})$
can be mapped to its depth image coordinates using the equation,
\begin{equation}
(x_{vc},y_{vc},z_{vc})^T=K_c(X_{vc}, Y_{vc}, Z_{vc})^T.
\label{c2ci}
\end{equation}
Based on the 2D coordinate $(x_{vc},y_{vc})$,
if voxel $v$ is outside the current depth camera frustum, the updating process
terminates and the algorithm moves on to update the next voxel.
If it is inside the current depth camera frustum, its actual depth value
$D_n(x_{vc},y_{vc})$ is
retrieved from the $n^{th}$ depth map $D_n$.

If $D_n(x_{vc},y_{vc})$ is non-zero, the valid depth map pixel
$(x_{vc},y_{vc}, D_n(x_{vc},y_{vc}))$
and its 3D point $(X_{vd}, Y_{vd}, Z_{vd})$ in the current camera coordinate system are
calculated as
\begin{equation}
(X_{vd}, Y_{vd}, Z_{vd})^T=K_c^{-1}(x_{vc},y_{vc}, D_n(x_{vc},y_{vc}))^T.
\label{vc2vd}
\end{equation}
If the Euclidean distance between $(X_{vc}, Y_{vc}, Z_{vc})$ and $(X_{vd},
Y_{vd}, Z_{vd})$
is greater than a threshold $\sigma$, the process terminates.
In our experiments, the threshold $\sigma$ is set to 20 (equivalently 20mm).

The weight of a point, $W$, is the dot product between the surface normal at
this point and the viewing direction from the camera to this point. The normal
vector for the current point $(X_{vd}, Y_{vd}, Z_{vd})$ is computed using
neighbouring retrojected points.
Assuming $(X_{1vd}, Y_{1vd}, Z_{1vd})$ and $(X_{2vd}, Y_{2vd}, Z_{2vd})$ are
the camera coordinates of $(x_{vc}+1,y_{vc}, D_n(x_{vc}+1,y_{vc})$ and
$(x_{vc},y_{vc}+1, D_n(x_{vc},y_{vc}+1)$, the normal is calculated using the
formula,
\begin{equation}
\begin{split}
N_v = &((X_{1vd}, Y_{1vd}, Z_{1vd})-(X_{vd}, Y_{vd}, Z_{vd})) \times \\
&((X_{2vd}, Y_{2vd}, Z_{2vd})-(X_{vd}, Y_{vd}, Z_{vd})).
\end{split}
\label{normal}
\end{equation}
If the weight, $W$, exceeds the previous weight times $0.8$ (i.e. $0.8 \cdot
W_{pre}$), the voxel is updated.
Since HD RGB frames are used to deliver high-resolution RGB images, for
colour voxel $v$ to be updated, its corresponding pixel in HD RGB image
$HD(x_{hd}, y_{hd})$ can be located considering camera intrinsic and extrinsic
parameters from RGB-D calibration (Section IV.$A$).
Then, the new colour $C_{new}$ in this colour voxel is updated as
\begin{equation}
C_{new}=\frac{C_{pre} \cdot W_{pre}+W \cdot HD(x_{hd}, y_{hd})}{W_{pre}+W}
\label{new_C}
\end{equation}
where $C_{pre}$ is the previous RGB value in this colour voxel and $HD(x_{hd},
y_{hd})$ is the RGB value of pixel $(x_{hd}, y_{hd})$ in the HD RGB image.
The new weight, $W_{new}$, is updated as,
\begin{equation}
W_{new}=W_{pre}+W \cdot (1-W_{pre})
\label{new_W}
\end{equation}
Finally, the new RGB value $C_{new}$ and the new weight $W_{new}$ are saved in the colour voxel $v$.

\subsection{Mesh Extraction and Simplification}

\begin{figure}[ht!]
\centering
\subfigure[][]{
\includegraphics[width=.25\linewidth]{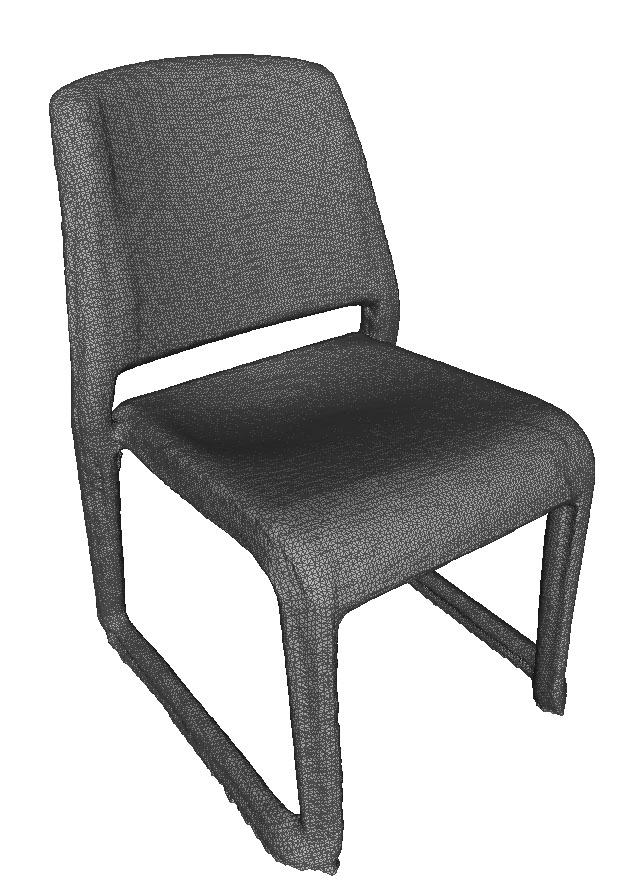}
\label{fig:ms1}}
\quad
\subfigure[][]{
\includegraphics[width=.25\linewidth]{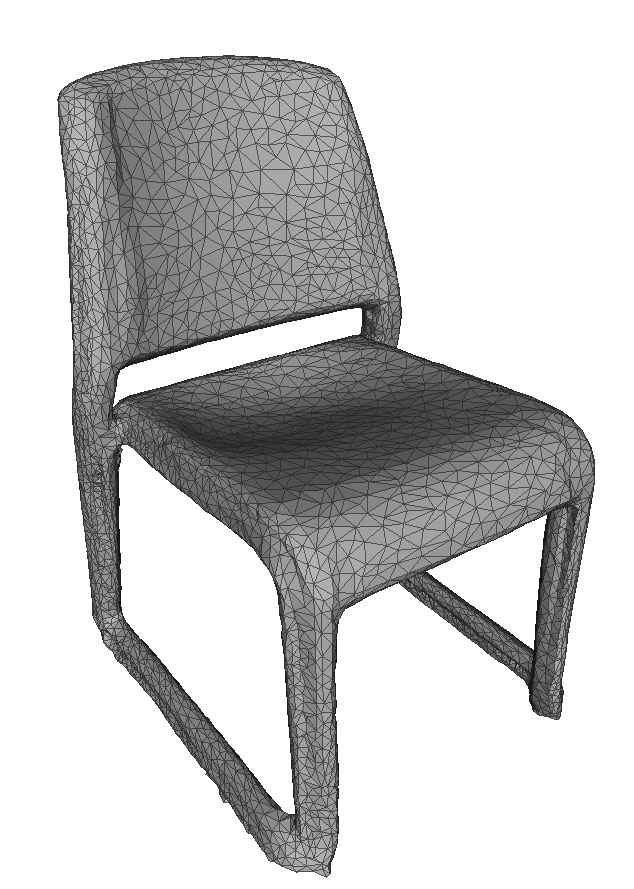}
\label{fig:ms2}}
\quad
\subfigure[][]{
\includegraphics[width=.25\linewidth]{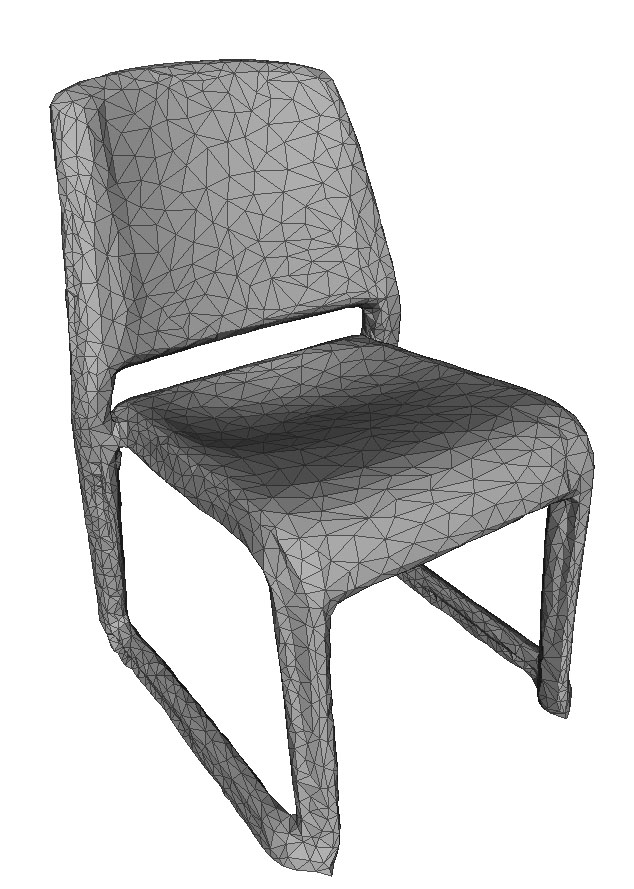}
\label{fig:ms3}}
\quad
\subfigure[][]{
\includegraphics[width=.25\linewidth]{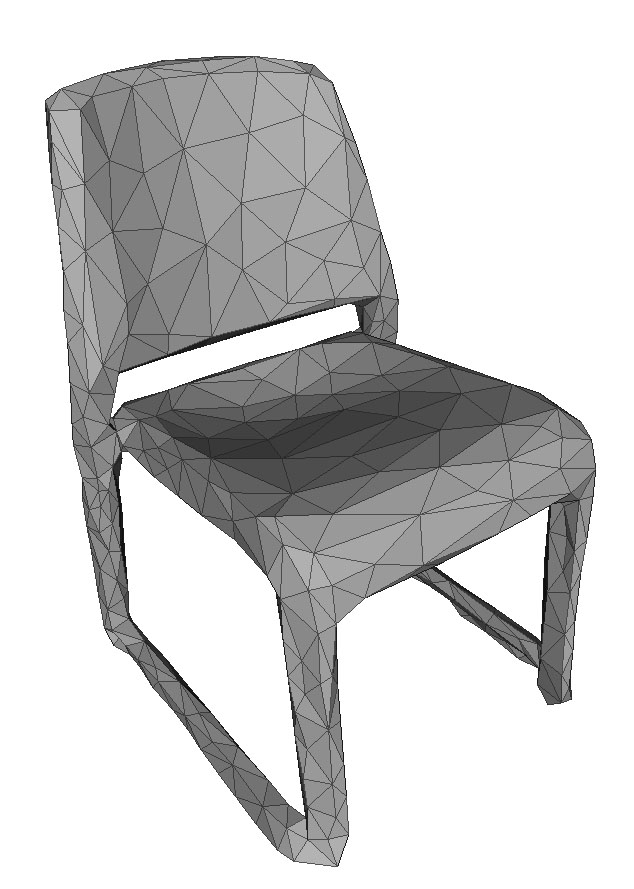}
\label{fig:ms4}}
\quad
\subfigure[][]{
\includegraphics[width=.25\linewidth]{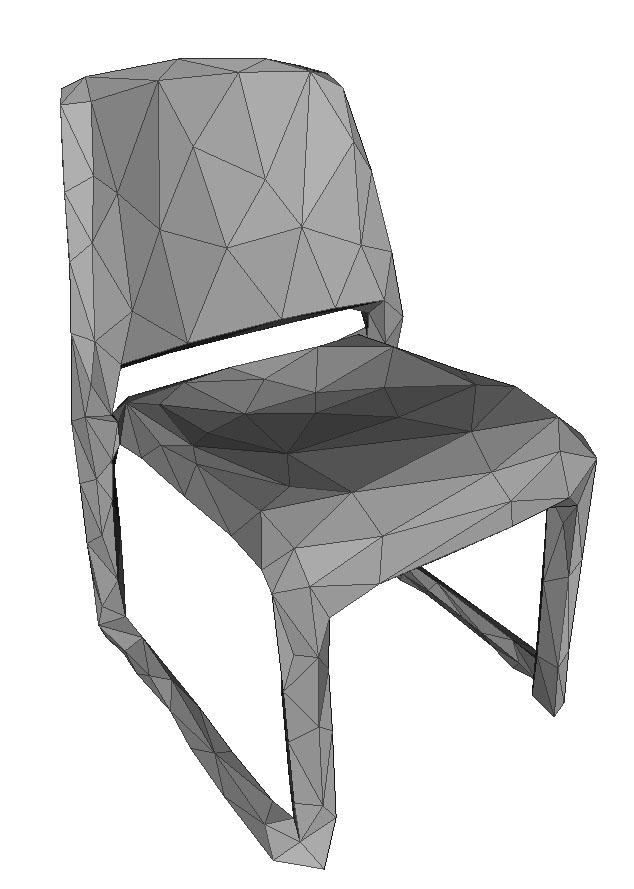}
\label{fig:ms5}}
\quad
\subfigure[][]{
\includegraphics[width=.25\linewidth]{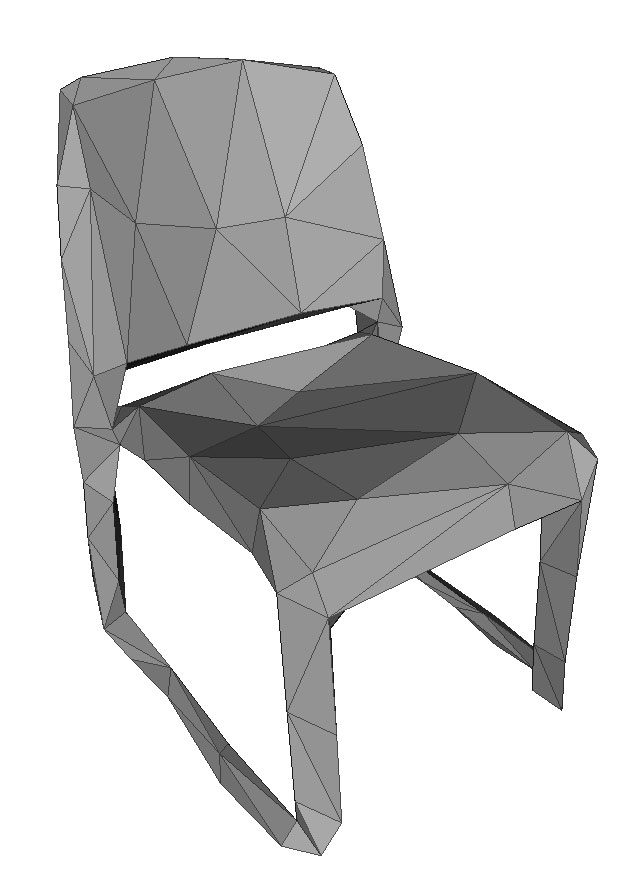}
\label{fig:ms6}}
\caption{Simplification of a chair model using quadric based mesh decimation.
(a): original dense model containing 137037 polygons.
(b): simplified model containing 13703 polygons (10\%).
(c): simplified model containing 6851 polygons (5\%).
(d): simplified model containing 1370 polygons (1\%)
(e): simplified model containing 685 polygons (0.5\%).
(f): simplified model containing 343 polygons (0.25\%).}
\label{fig:model_simplify}
\end{figure}

The marching cube algorithm is used to extract 3D polygon mesh models from TSDF
volume in KinectFusion. However, dense polygon mesh models obtained directly
from marching cube algorithm tend to contain a large number of redundant
polygons, which can increase the model
complexity without providing much geometry details.
When memory and rendering efficiency are considered, simplification of the
reconstructed dense polygon mesh is necessary in many circumstances.
In order to achieve model complexity reduction the quadric-based mesh
decimation is adopted\cite{Garland:1997:SSU:258734.258849}.
Fig. \ref{fig:model_simplify} shows the results of simplifying a dense chair
model.
The original dense chair model (Fig. \ref{fig:ms1}) is decimated to contain
different numbers of polygons.
It can be observed that all simplified models (see Fig.~\ref{fig:ms2} -
Fig.~\ref{fig:ms6}) can still keep their shape as a chair.
The results indicate that quadric-based mesh decimation algorithm is effective
in simplifying polygon mesh model while preserving geometric features.

In passing, note that the reduction rate parameter of the decimation
algorithm specifies the degree to which the current mesh model will be
decimated. For instance, reduction rate of 10\% implies that the simplified
mesh model after polygon reduction will contain 10\% of polygons as the original
model. This reduction rate can be adjusted according to the actual
requirements of 3D model generated.

\subsection{Texture Generation}

\subsubsection{2D Texture Map Generation for Each Polygon}

Compared with a dense polygon mesh, a simplified model contains much less polygons.
Texturing by coloured vertex method will incur losses of texture details
which in turn reduces the texture quality.
To maintain texture quality on a model with less polygons, each polygon
should be textured with more RGB details. To this end, different 2D texture maps
are required to be generated and mapped to all polygons in a mesh model; one 2D
texture map for one polygon.

\begin{figure}[h]
\centering
\subfigure[][]{
\includegraphics[width=.3\linewidth]{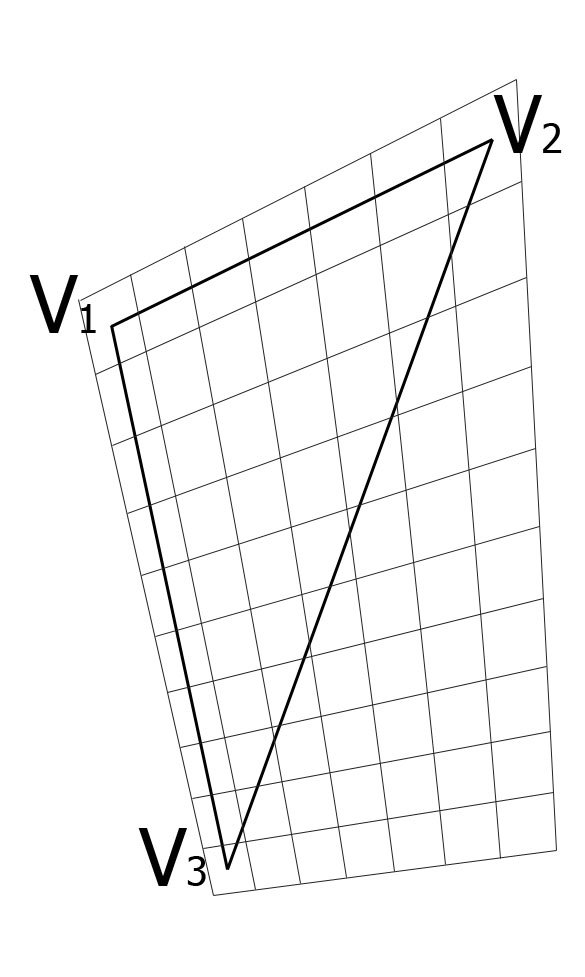}
\label{tex_cord:l}}
\subfigure[][]{
\includegraphics[width=.55\linewidth]{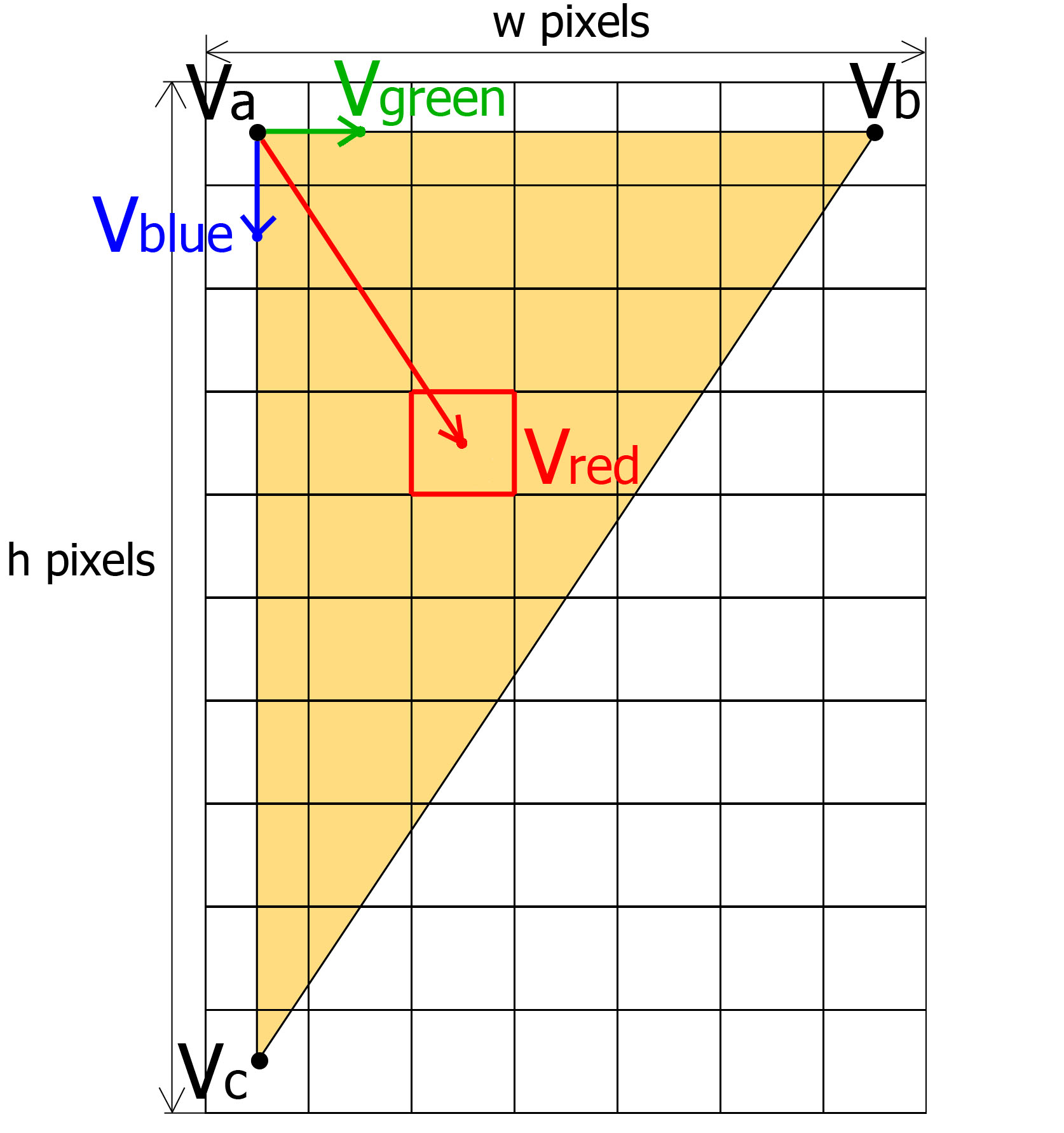}
\label{tex_cord:r}}
\caption{Generating 2D texture map, each block represents one pixel in the 2D texture map.
(a): a polygon in the model to be textured.
(b): a 2D texture map for the polygon in (a).}
\label{tex_cord}
\end{figure}

In the reconstructed 3D models, each polygon is a triangle with 3 vertices.
For an arbitrary polygon in a 3D model, assume its
3 vertices are $(V_1, V_2, V_3)$ (Fig \ref{tex_cord:l}); $V_1, V_2, V_3$
are points with 3D space coordinates in real world.
When generating a 2D texture map (Fig \ref{tex_cord:r}) for
this particular polygon, the upper left triangle in the texture map
($\bigtriangleup V_aV_bV_c$ in Fig \ref{tex_cord:r}) will be mapped onto this
polygon.
The mapping relations are defined as follows: $V_a \mapsto V_1$; $V_b
\mapsto V_2$; and  $V_c \mapsto V_3$

The resolution of the texture image is calculated as the actual
size of polygon divided by a given pixel size parameter $Size_{pixel}$.
The value of $Size_{pixel}$ represents the size of a pixel when
mapped onto a 3D model, which should be set according to the HD RGB camera
resolution and scanning distance.
In Fig.~\ref{tex_cord:r}, the resolution of 2D texture map is $w \times h$
pixels, where
\begin{equation}
w=\text{distance}(V_a, V_b)/Size_{pixel}
\label{wv1v2}
\end{equation}
\begin{equation}
h=\text{distance}(V_a, V_c)/Size_{pixel}
\label{hv1v3}
\end{equation}
$\text{distance}(V_i, V_j)$ is the Euclidean distance between 3D coordinates
$V_i$ and $V_j$.

In order to complete the texture generation process, two unit vectors
($\overrightarrow{V_a V_{green}}, \overrightarrow{V_a V_{blue}}$) are
required;
\begin{equation}
\overrightarrow{V_a V_{green}}=\overrightarrow{V_a V_b} / (w-1),
\label{v_green}
\end{equation}
\begin{equation}
\overrightarrow{V_a V_{blue}}=\overrightarrow{V_a V_c} / (h-1).
\label{v_blue}
\end{equation}

For any pixel in the texture map, its 3D world
position can be expressed as a combination of these two vectors.
If the pixel marked with a red bounding box is considered for
example, its 3D world position can be expressed as $V_a + \overrightarrow{V_a
V_{red}}$. This is equivalent to $V_a + 2 \overrightarrow{V_a
V_{green}} + 3
\overrightarrow{V_a V_{blue}}$.
Given the 3D world positions, the RGB value of any
pixel can be found by accessing the colour volume.
By repeating the process above for every pixel in
the texture map, a complete 2D texture map can be obtained.

\subsubsection{2D Texture Map Merging}

A regular 3D model (including simplified ones) usually contains hundreds
or thousands of polygons.
The generation of 2D textures for each polygon will result in the production of
very large number of 2D texture maps.
Loading each texture map one at a time is time consuming and required a speedup
strategy. An efficient method is to load multiple textures onto one or a few
2D images. This strategy is implemented in 2D texture merging
scheme.

2D texture image that are supposed to contain texture maps for multiple polygons are created.
Each texture map for one polygon is added and placed in the 2D image starting
from the top-left corner to bottom-right corner in a column-wise order.

\begin{figure}[h]
\centering
\subfigure[][]{
\includegraphics[width=.4\linewidth]{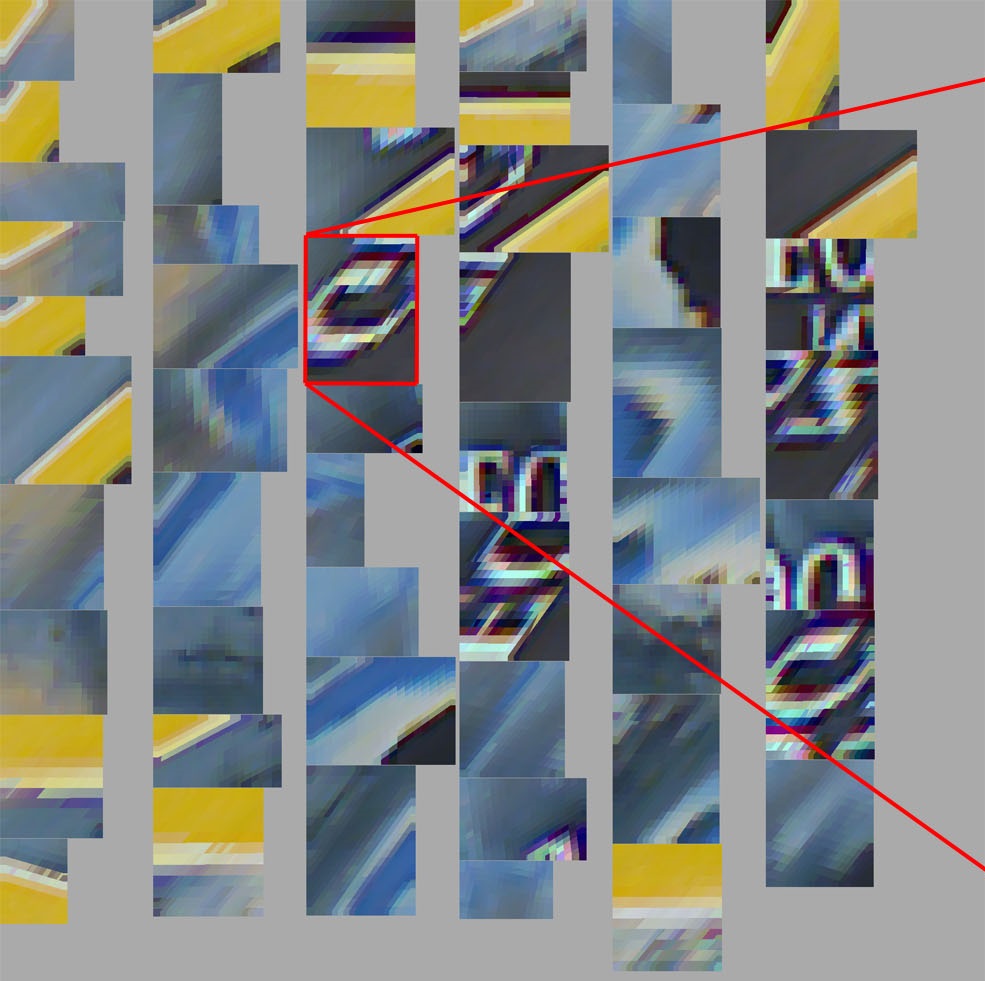}
\label{fig:texture1:l}}
\subfigure[][]{
\includegraphics[width=.33\linewidth]{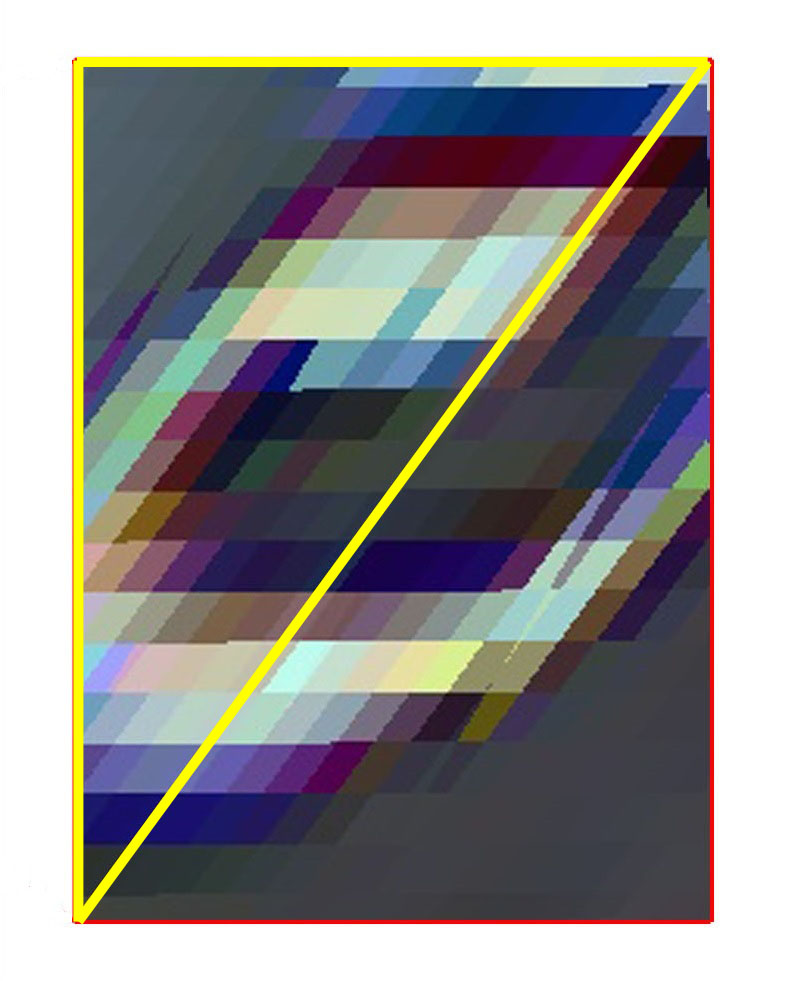}
\label{fig:texture1:r}}
\caption{
(a): a merged 2D texture image file.
(b): a 2D texture map for one polygon. Best viewed in colour.}
\label{fig:texture1}
\end{figure}

Fig.~\ref{fig:texture1} depicts one 2D texture image file created by merging
multiple texture maps.
Fig.~\ref{fig:texture1:l} shows the merged texture image and
Fig.~\ref{fig:texture1:r} is a 2D texture map for one polygon.
The upper left triangle of each 2D texture map (yellow triangle in Fig
\ref{fig:texture1:r}) will be rendered on a polygon as the actual model texture.

\section{Experiments and Results}

The HD RGB camera used in our experiment is a Point Grey GS3-U3-41C6C-C camera with an 8mm F1.4 wide angle lens.
Our 3D reconstruction method is implemented in C++ and run on a PC with a Nvidia GTX680 GPU, 4GB graphical memory.

Experiments are carried out on 2 test cases to verify the effectiveness of our texturing method in terms of
\begin{itemize}
\item texture improvements from using 2D texture mapping compared with using coloured vertex; and
\item texture improvements from using HD RGB images compared with using Kinect RGB images.
\end{itemize}

The test cases selected include a lunch bag and a backpack, which are small and
medium-sized objects with textures of noticeable details (e.g. brand logo).
In our experiment, the dimension of geometry volume is $384\times384\times384$
and the asymmetrical colour volume is set to $768\times768\times768$.
Volume size is variable to suit the size of the object to be reconstructed.

To obtain a visual evaluation of texture quality, reconstructed 3D models with
textures are loaded and visualized in MeshLab. A snapshot from a fixed view
point is taken for each model.
An objectve evaluation of our algorithm measures the level of texture quality.
For each test case, image patches are cropped from the same location where
patterns containing detailed texture could be found (e.g. brand logo).
An image patch from high quality textures tend to be sharp, so that more
texture details can be captured.
Therefore, image sharpness of the image patch is measured as the indicator of
texture quality using gradient magnitude; that is, the sum of all
gradient norms divided by number of pixels.
A higher gradient magnitude implies more sharpness and more texture details.

\subsection{RGB-D Camera Calibration Results}

Efforts to ensure correct RGB texture mapping require an accurate calibration
between HD RGB camera and the Kinect depth camera. An evaluation
on RGB-D camera calibration is performed before our experiments.

In the process of RGB-D calibration, 100 sample shots were taken which include
100 HD RGB images and 100 depth maps.
These samples can be further divided into 3 groups according to the average
shooting distance (distance between the camera and the chess board pattern),
which include short range(0.5M-1M), middle range(1M-2.5M) and long
range(2.5M-4M). In each group, there are more than 30 RGB-D image pairs that
were taken from different view points of the chess board.
RGB-D calibration is then performed using all sample shots.

\begin{figure}[ht]
\centering
\subfigure[][]{
\includegraphics[width=.4\linewidth]{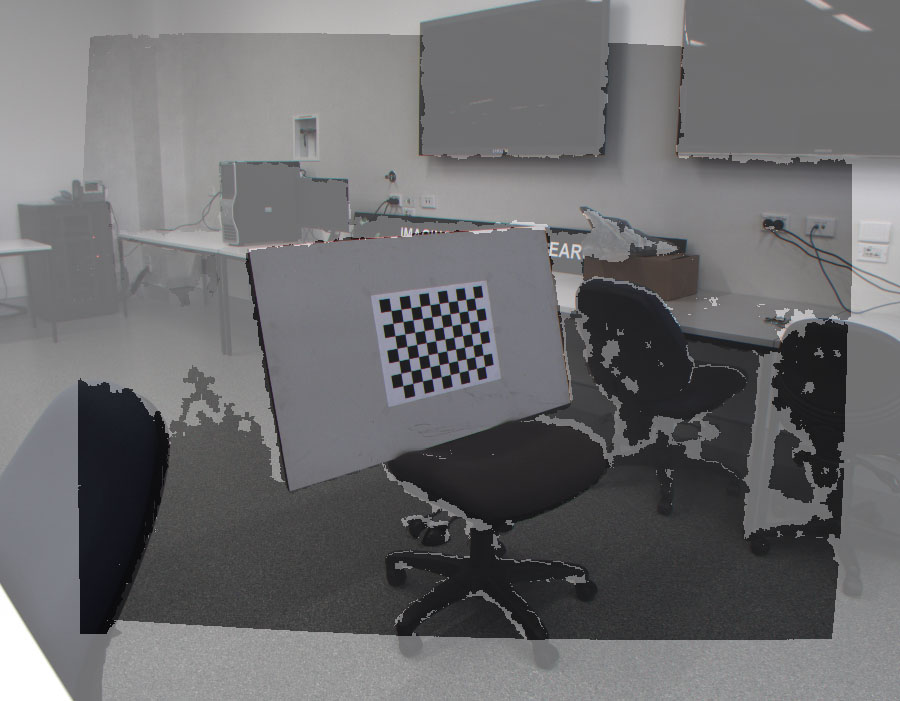}
\label{fig:calib_res_1}}
\subfigure[][]{
\includegraphics[width=.4\linewidth]{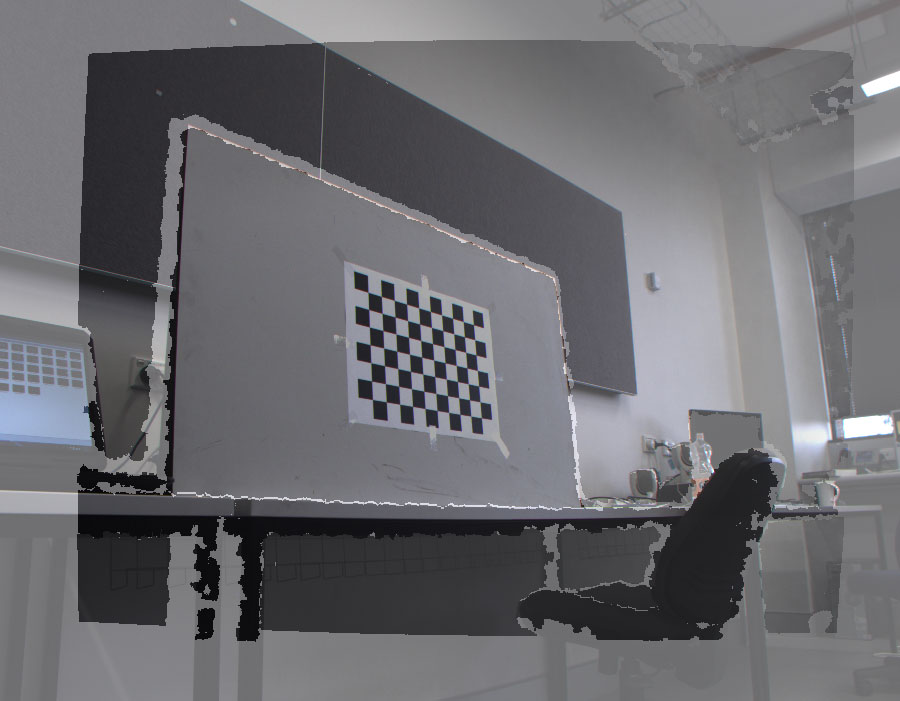}
\label{fig:calib_res_2}}
\caption{Overlapped HD RGB image with registered depth map}
\label{reg_res}
\end{figure}

The accuracy of RGB-D camera calibration is ascertained a straightforward way
by visually checking the overlapped HD RGB image and visualised depth map for
misalignment. The overlayed image can be generated by overlaying the aligned
semitransparent depth map onto its corresponding HD RGB image
(Fig.~\ref{reg_res}). Judging by the overlayed images, it can be observed that
the depth map and HD RGB images are reasonably well aligned.

\subsection{3D Model Texture Results}

\subsubsection{Improvements from using 2D texture mapping}

Most Kinectfusion-based methods texture the reconstructed models using coloured
vertex; assignment of an RGB value to each vertex in the model.
The texture of a polygon is the linear interpolation of 3 RGB value on its
vertices, and this can lead to loss of details due to mesh simplification.
On the other hand, our proposed texture method generate 2D texture maps
directly from the colour volume and this has maintained texture details even on
a simplified 3D polygon mesh.

\begin{figure*}[p]
\centering
\subfigure[][]{
\includegraphics[width=.14\linewidth]{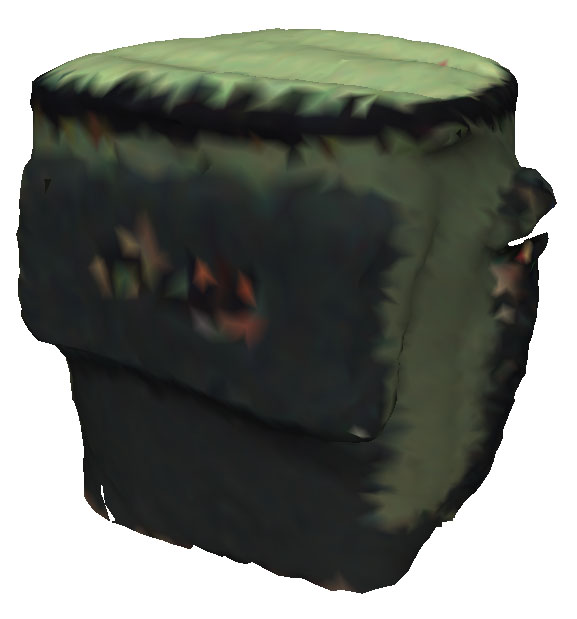}
\label{fig:res:bbp1}}
\subfigure[][]{
\includegraphics[width=.14\linewidth]{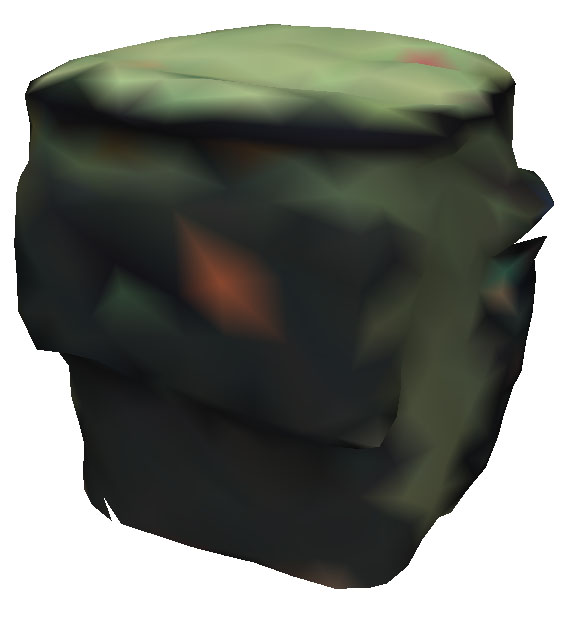}
\label{fig:res:bbp2}}
\subfigure[][]{
\includegraphics[width=.14\linewidth]{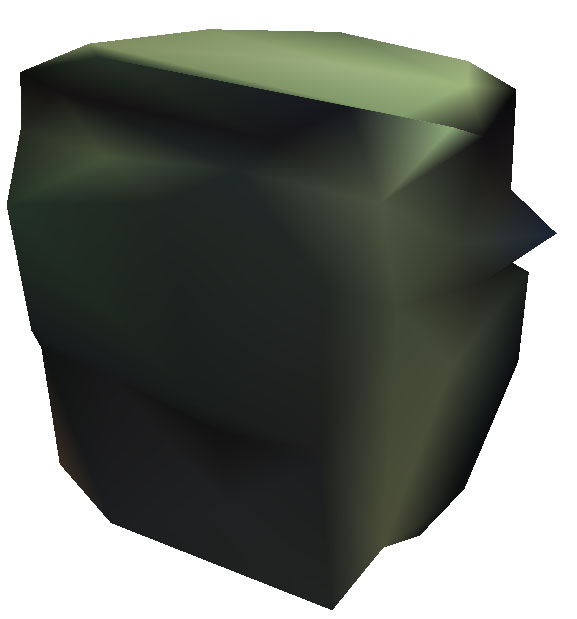}
\label{fig:res:bbp3}}
\subfigure[][]{
\includegraphics[width=.14\linewidth]{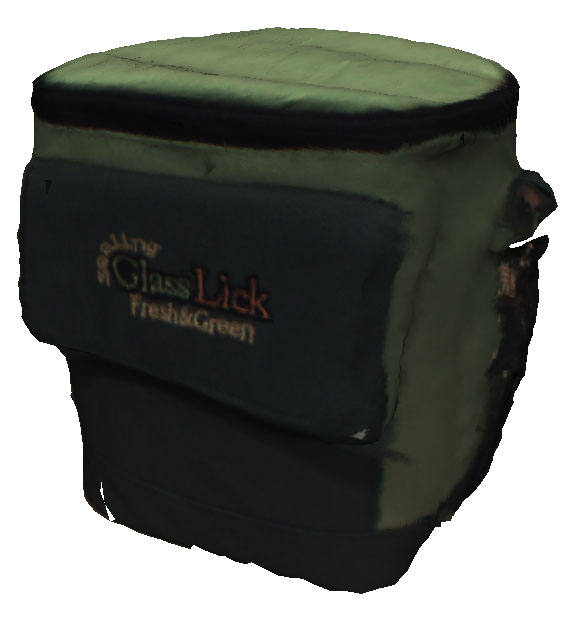}
\label{fig:res:bbt1}}
\subfigure[][]{
\includegraphics[width=.14\linewidth]{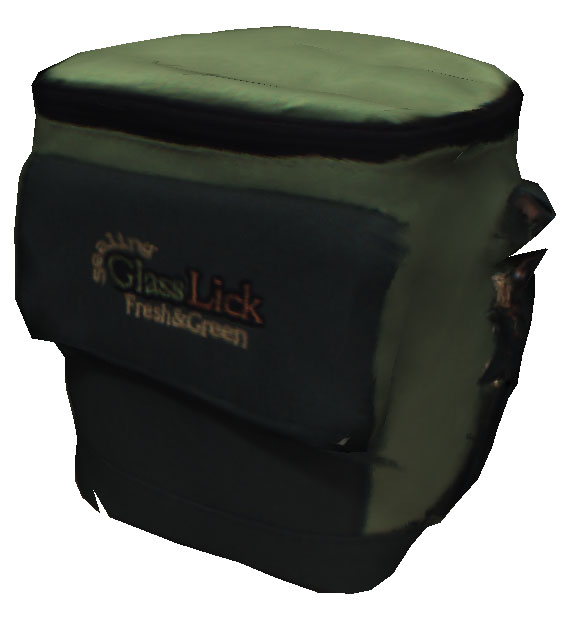}
\label{fig:res:bbt2}}
\subfigure[][]{
\includegraphics[width=.14\linewidth]{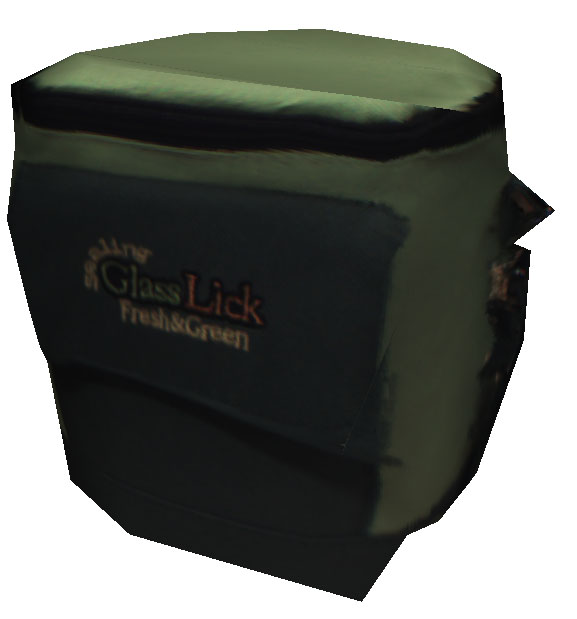}
\label{fig:res:bbt3}}
\subfigure[][]{
\includegraphics[width=.14\linewidth]{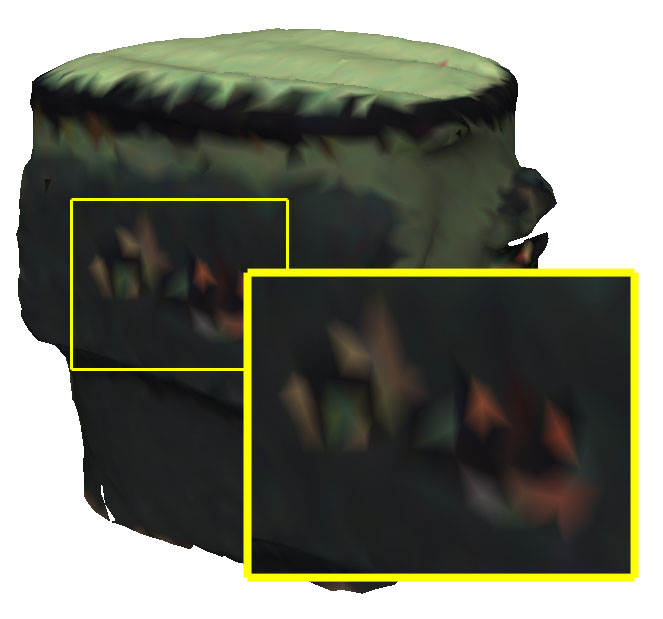}
\label{fig:res:bb_wp_cv_1}}
\subfigure[][]{
\includegraphics[width=.14\linewidth]{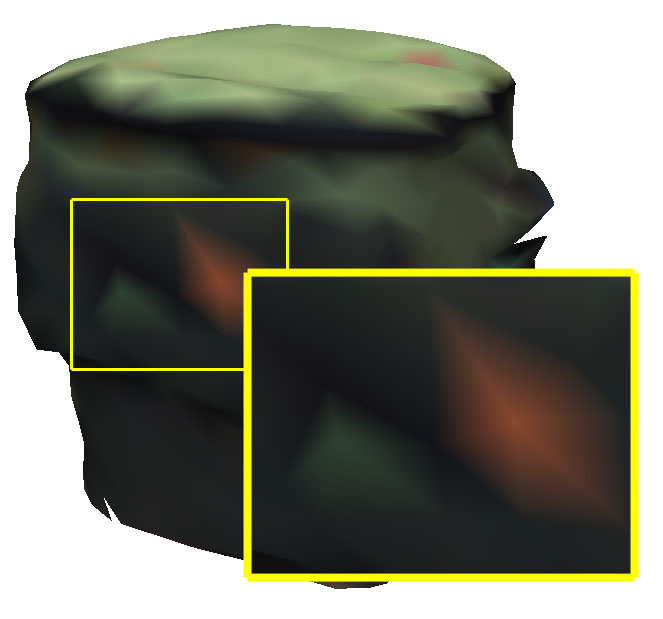}
\label{fig:res:bb_wp_cv_2}}
\subfigure[][]{
\includegraphics[width=.14\linewidth]{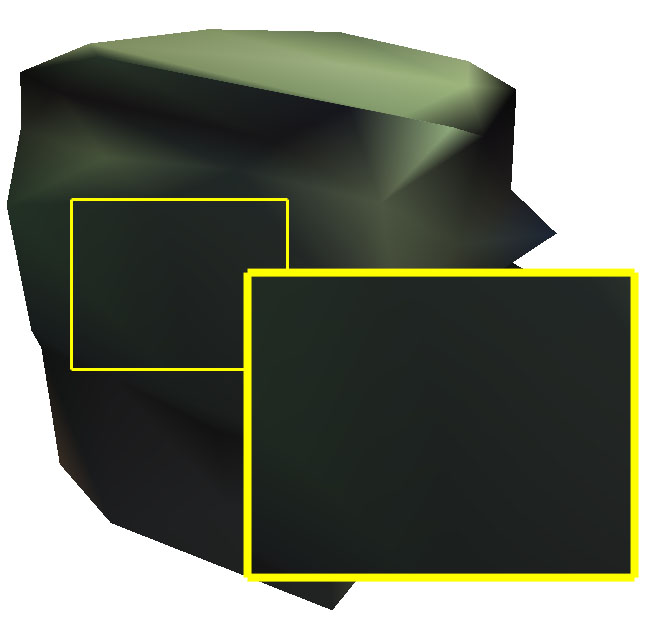}
\label{fig:res:bb_wp_cv_3}}
\subfigure[][]{
\includegraphics[width=.14\linewidth]{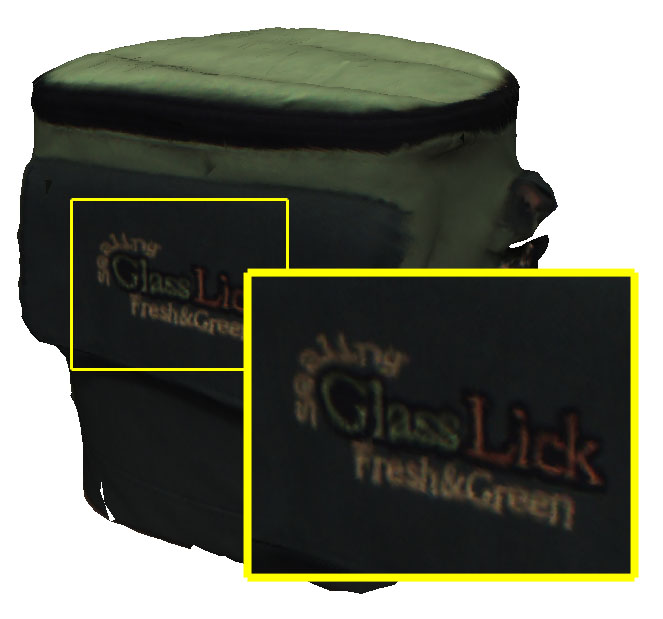}
\label{fig:res:bb_wp_2d_1}}
\subfigure[][]{
\includegraphics[width=.14\linewidth]{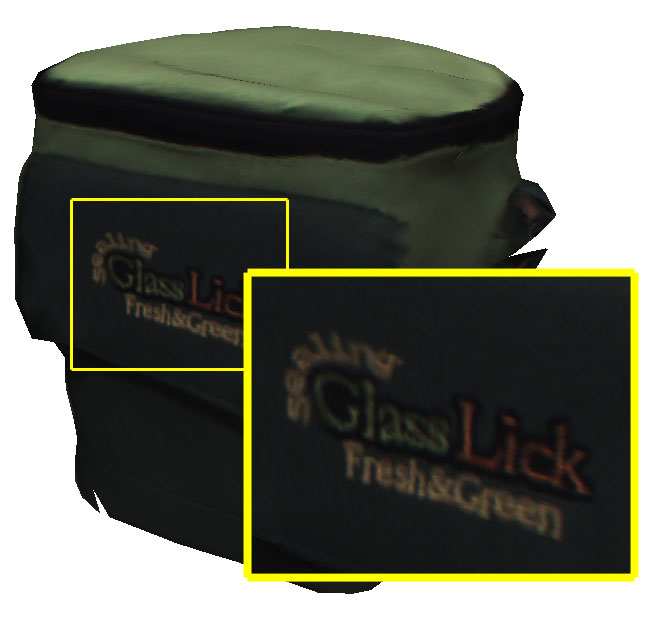}
\label{fig:res:bb_wp_2d_2}}
\subfigure[][]{
\includegraphics[width=.14\linewidth]{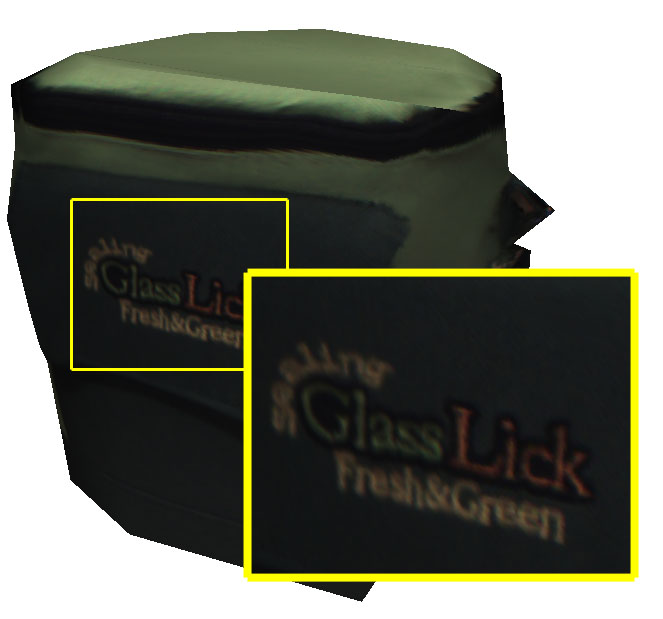}
\label{fig:res:bb_wp_2d_3}}
\caption{Texture results of a lunch bag model
(a): coloured vertex model with 13360 polygons.
(b): coloured vertex model with 1336 polygons.
(c): coloured vertex model with 134 polygons.
(d): 2D textured model with 13360 polygons.
(e): 2D textured model with 1336 polygons.
(f): 2D textured model with 134 polygons.
(g): image patch of (a).
(h): image patch of (b).
(i): image patch of (c).
(j): image patch of (d).
(k): image patch of (e).
(l): image patch of (f).}
\label{fig:results:bb}
\end{figure*}

\begin{figure*}[p]
\centering
\subfigure[][]{
\includegraphics[width=.14\linewidth]{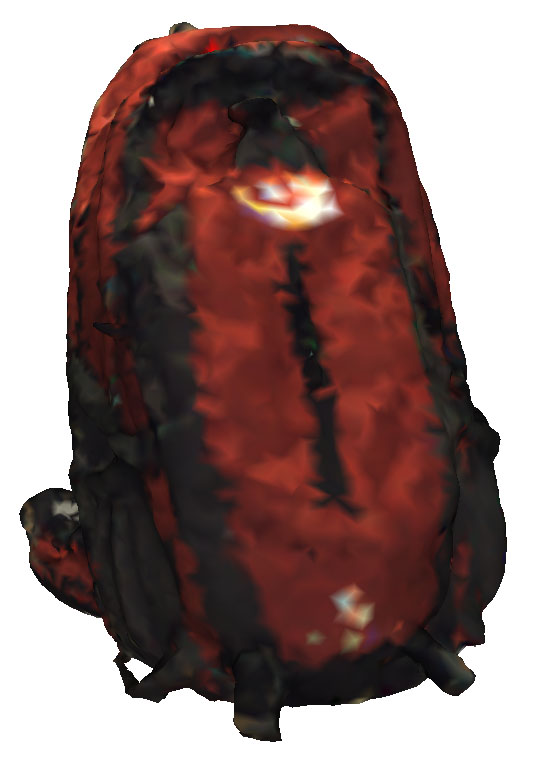}
\label{fig:res:b2p0}}
\subfigure[][]{
\includegraphics[width=.14\linewidth]{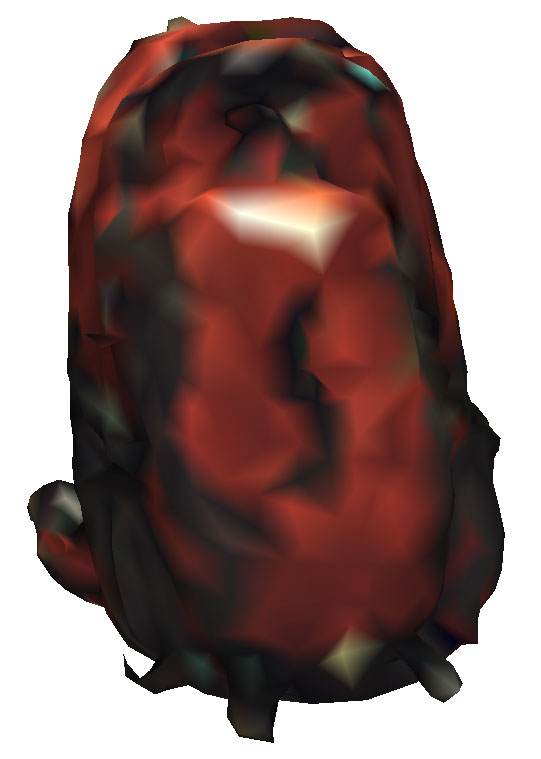}
\label{fig:res:b2p1}}
\subfigure[][]{
\includegraphics[width=.14\linewidth]{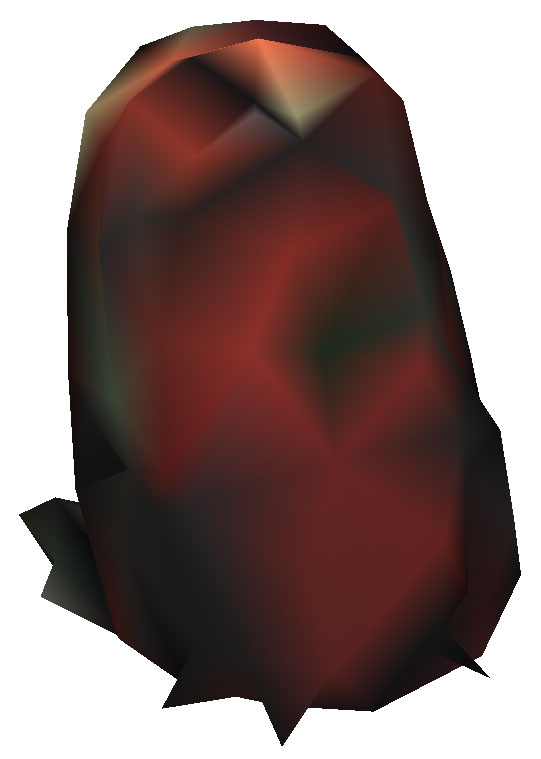}
\label{fig:res:b2p2}}
\subfigure[][]{
\includegraphics[width=.14\linewidth]{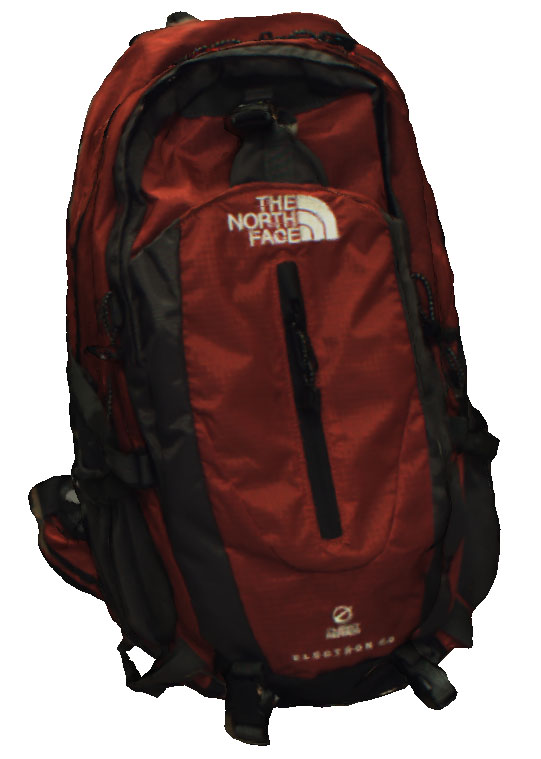}
\label{fig:res:b2t0}}
\subfigure[][]{
\includegraphics[width=.14\linewidth]{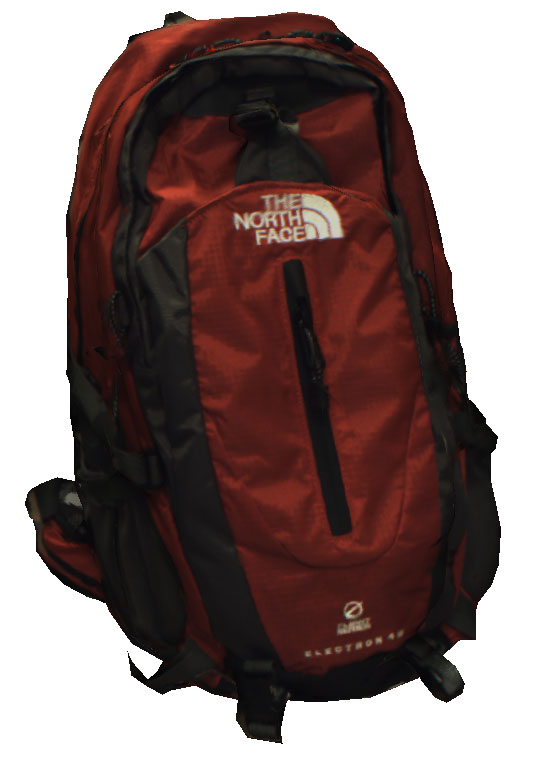}
\label{fig:res:b2t1}}
\subfigure[][]{
\includegraphics[width=.14\linewidth]{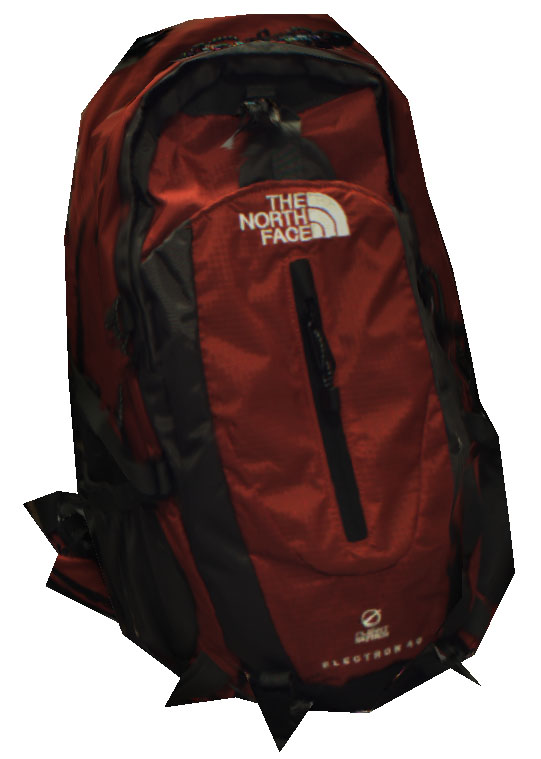}
\label{fig:res:b2t2}}
\subfigure[][]{
\includegraphics[width=.14\linewidth]{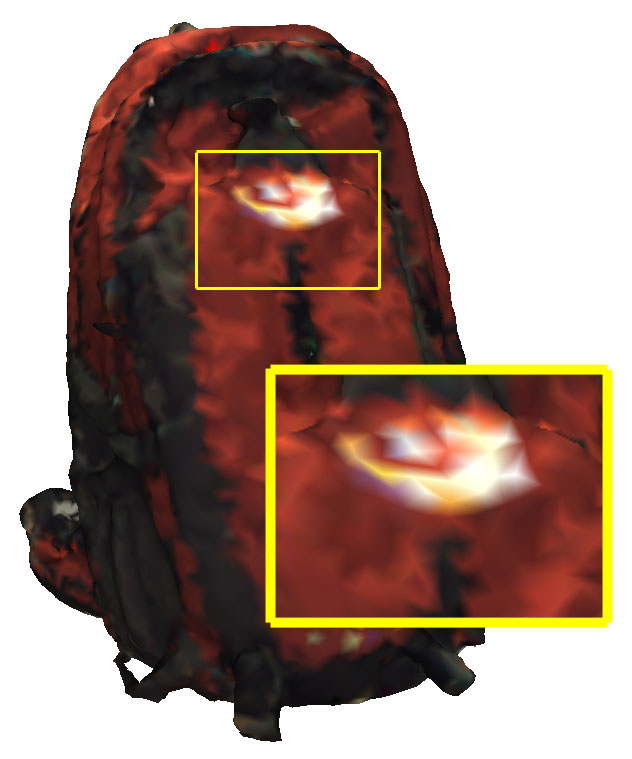}
\label{fig:res:b2_wp_cv_1}}
\subfigure[][]{
\includegraphics[width=.14\linewidth]{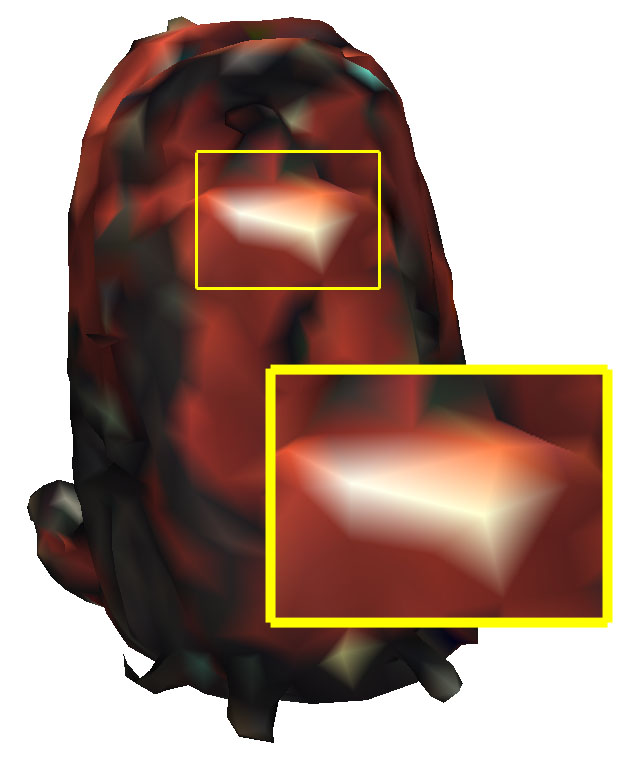}
\label{fig:res:b2_wp_cv_2}}
\subfigure[][]{
\includegraphics[width=.14\linewidth]{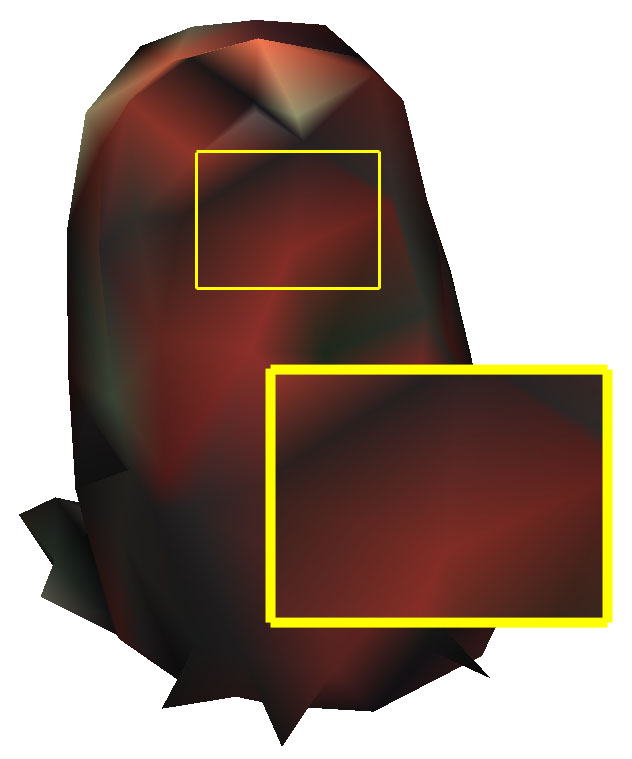}
\label{fig:res:b2_wp_cv_3}}
\subfigure[][]{
\includegraphics[width=.14\linewidth]{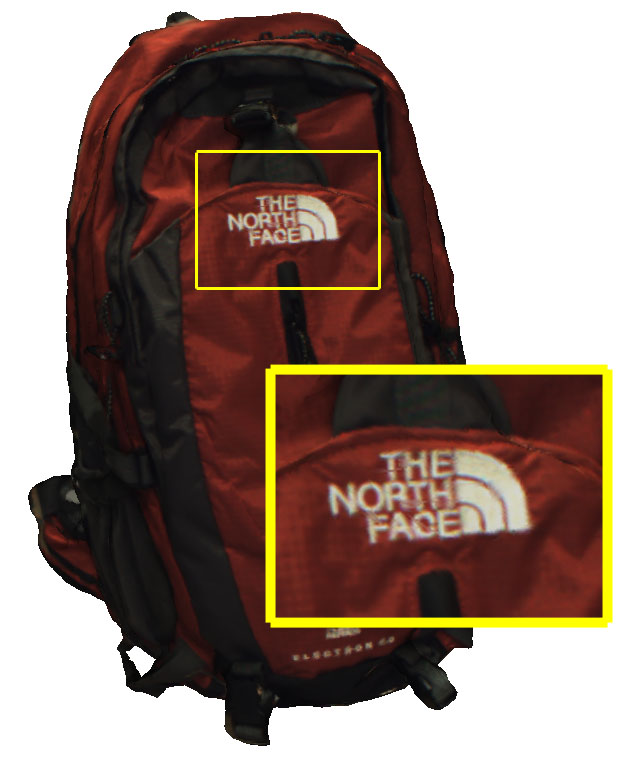}
\label{fig:res:b2_wp_2d_1}}
\subfigure[][]{
\includegraphics[width=.14\linewidth]{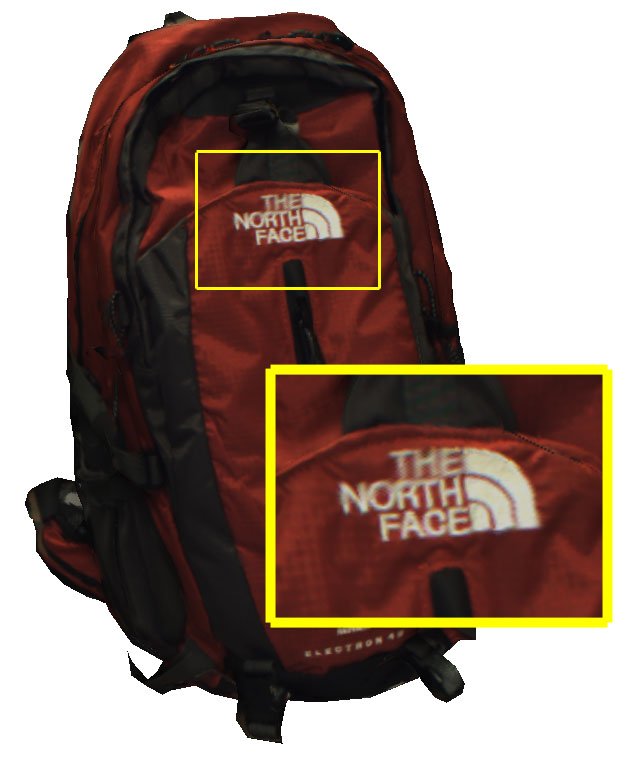}
\label{fig:res:b2_wp_2d_2}}
\subfigure[][]{
\includegraphics[width=.14\linewidth]{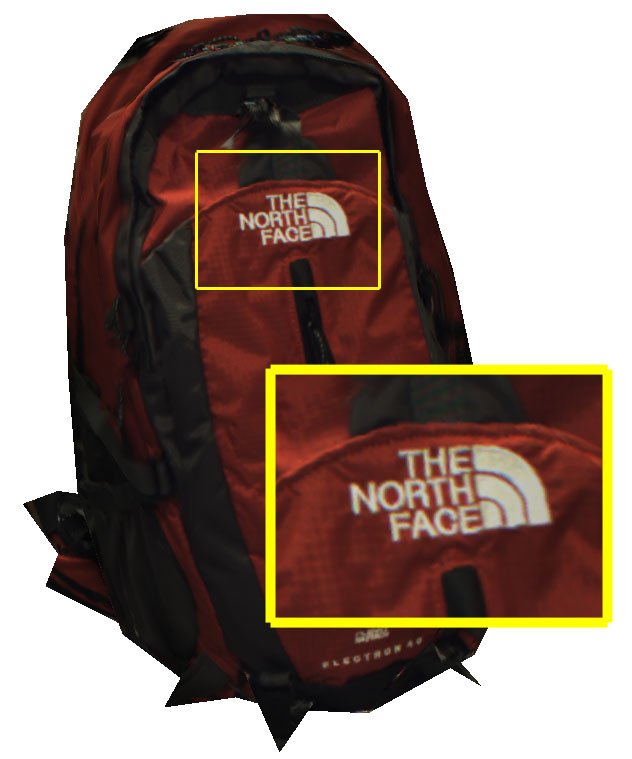}
\label{fig:res:b2_wp_2d_3}}
\caption{Texture results of a backpack model
(a): coloured vertex model with 39176 polygons.
(b): coloured vertex model with 3918 polygons.
(c): coloured vertex model with 392 polygons.
(d): 2D textured model with 39176 polygons.
(e): 2D textured model with 3918 polygons.
(f): 2D textured model with 392 polygons.
(g): image patch of (a).
(h): image patch of (b).
(i): image patch of (c).
(j): image patch of (d).
(k): image patch of (e).
(l): image patch of (f).}
\label{fig:results:b2}
\end{figure*}

Fig.~\ref{fig:results:bb} and Fig \ref{fig:results:b2} depict simplified models
that are textured with these two methods.
Table~\ref{tab:vcv2d} presents texture qualities of reconstructed models
as measured by gradient magnitudes of their image patches.
From the table, the quality of texture generated using coloured vertex drops
drastically as the level of model simplification increases.
However, our 2D textured mapping method is able to maintain the texture quality
at a relatively high level with only a small texture degradation.

Assessments based on visual check also support the texture results shown in
Table \ref{tab:vcv2d}. It can be observed that texture details on coloured
vertex models are greatly ``washed'' away by mesh simplification.
Texture details such as letters and brand logos are difficult to distinguish
on those models.
On the other hand, the letters and brand logos are clear to see and easy
to distinguish from models textured using our 2D texture mapping.

Based on the results shown, it can be verified that our proposed texture method
is effective in terms of preserving texture qualities on a simplified polygon
mesh model.

\begin{table}[ht]
\small
\centering
\caption{Texture qualities (coloured vertex vs. 2D texture mapping)}
\label{tab:vcv2d}
\begin{tabular}{|c|c|c|c|}
\hline
{\bf \begin{tabular}[c]{@{}c@{}}Image\\ Patch\end{tabular}} & {\bf \begin{tabular}[c]{@{}c@{}}Gradient\\ Magnitude\end{tabular}} & {\bf \begin{tabular}[c]{@{}c@{}}Image\\ Patch\end{tabular}} & {\bf \begin{tabular}[c]{@{}c@{}}Gradient\\ Magnitude\end{tabular}} \\ \hline
\multicolumn{4}{|c|}{{\bf Test Case: Lunch Bag}}                                                                                                                                                                                                                    \\ \hline
\multicolumn{2}{|c|}{{\bf Coloured Vertex}}                                                                                      & \multicolumn{2}{c|}{{\bf 2D Texture Mapping}}                                                                                    \\ \hline
Fig \ref{fig:res:bb_wp_cv_1}                                                           & 1.1292                                                                  & Fig \ref{fig:res:bb_wp_2d_1}                                                           & 2.1064                                                                  \\ \hline
Fig \ref{fig:res:bb_wp_cv_2}                                                           & 0.6633                                                                  & Fig \ref{fig:res:bb_wp_2d_2}                                                           & 1.9707                                                                  \\ \hline
Fig \ref{fig:res:bb_wp_cv_3}                                                           & 0.0870                                                                  & Fig \ref{fig:res:bb_wp_2d_3}                                                           & 1.8888                                                                  \\ \hline
\multicolumn{4}{|c|}{{\bf Test Case: Backpack}}                                                                                                                                                                                                                     \\ \hline
\multicolumn{2}{|c|}{{\bf Coloured Vertex}}                                                                                      & \multicolumn{2}{c|}{{\bf 2D Texture Mapping}}                                                                                    \\ \hline
Fig \ref{fig:res:b2_wp_cv_1}                                                           & 3.9389                                                                  & Fig \ref{fig:res:b2_wp_2d_1}                                                           & 5.7487                                                                  \\ \hline
Fig \ref{fig:res:b2_wp_cv_2}                                                           & 2.6984                                                                  & Fig \ref{fig:res:b2_wp_2d_2}                                                           & 5.4861                                                                  \\ \hline
Fig \ref{fig:res:b2_wp_cv_3}                                                           & 0.4366                                                                  & Fig \ref{fig:res:b2_wp_2d_3}                                                           & 5.3571                                                                  \\ \hline
\end{tabular}
\normalsize
\end{table}


\subsubsection{Improvements from using HD RGB Image}

HD RGB camera can deliver RGB images with a higher resolution than that from
the Kinect RGB camera. Thus such a camera is able to capture more details of
real world.
Based on this observation, model texture generated by using HD RGB image
should also contain more texture details.
To verify this assumption, the following experiment is carried out.

For each test case, a model textured using Kinect RGB image and a model
textured using HD RGB image are generated and compared.
The geometry of these two models are exactly the same and both models are
textured using 2D texture mapping method to keep high texture qualities.

\begin{figure}[ht]
\centering
\subfigure[][]{
\includegraphics[width=.4\linewidth]{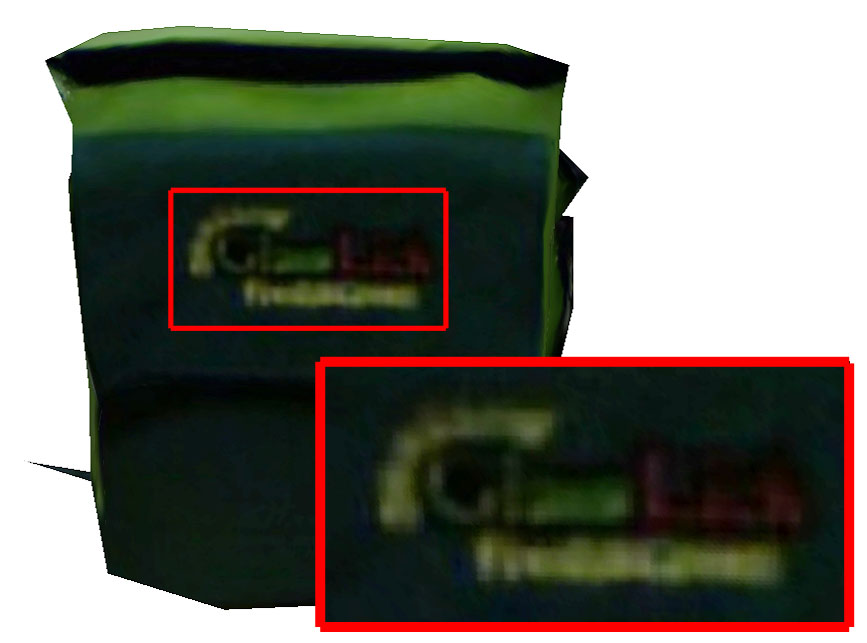}
\label{fig:bb_texture_zoom_k}}
\subfigure[][]{
\includegraphics[width=.4\linewidth]{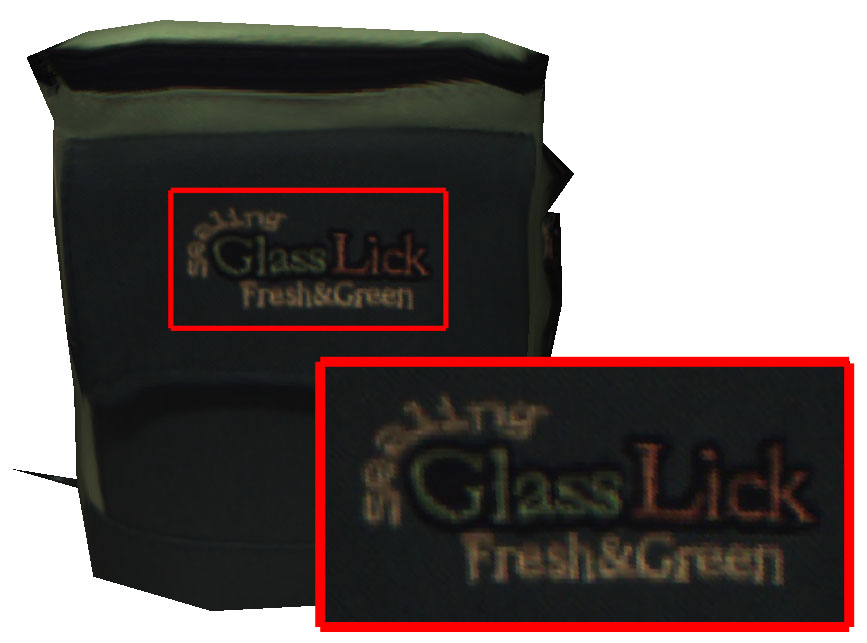}
\label{fig:bb_texture_zoom_c}}
\caption{Textured model and cropped image patch of a lunch bag
(a): texture using Kinect RGB.
(b): texture using HD RGB.}
\label{fig:bb_kvhd}
\end{figure}

\begin{figure}[ht]
\centering
\subfigure[][]{
\includegraphics[width=.4\linewidth]{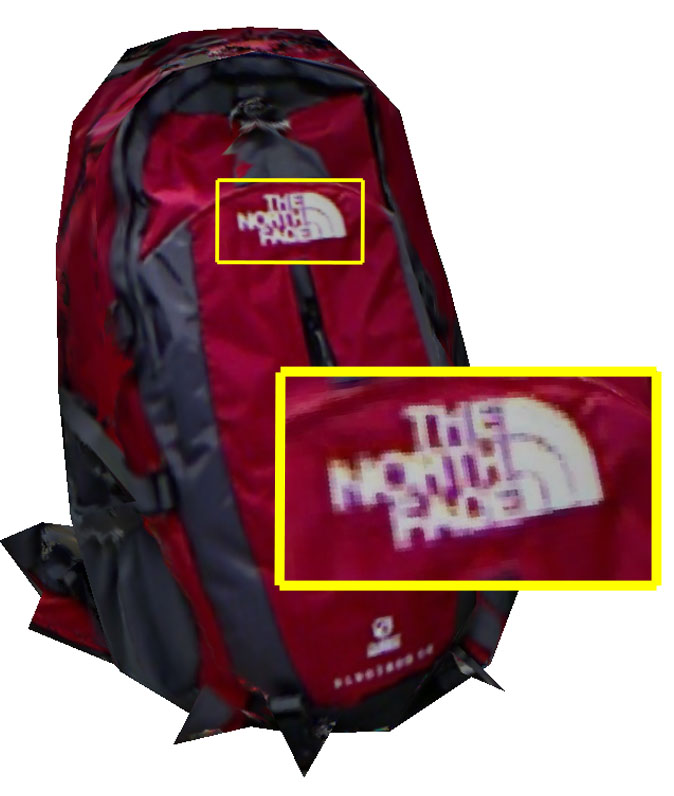}
\label{fig:b2_texture_zoom_c_k}}
\subfigure[][]{
\includegraphics[width=.4\linewidth]{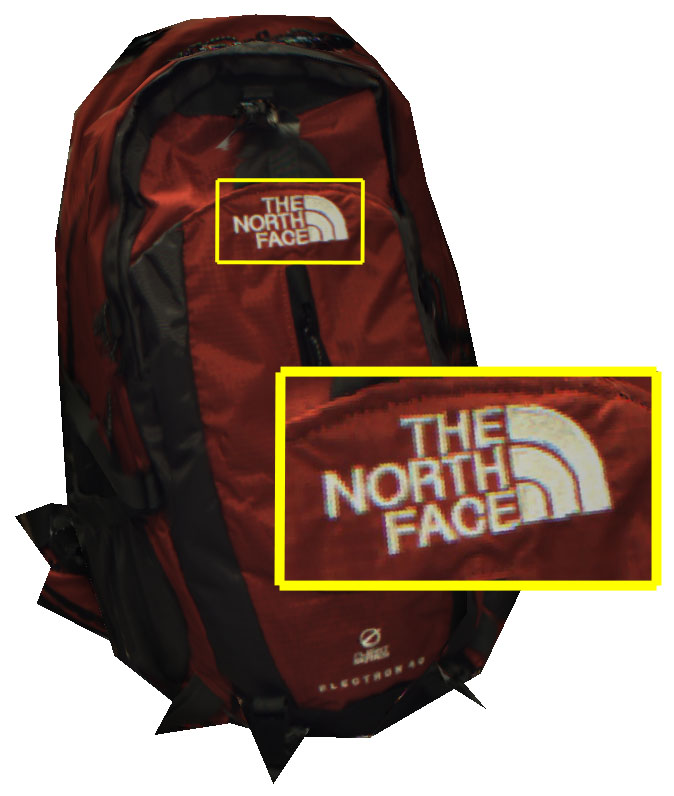}
\label{fig:b2_texture_zoom_c}}
\caption{Textured model and cropped image patch of a backpack
(a): texture using Kinect RGB.
(b): texture using HD RGB.}
\label{fig:b2_kvhd}
\end{figure}

Table~\ref{tab:kvhd} shows the difference of texture quality between using
Kinect RGB and HD RGB. For both test cases, textures generated by using HD RGB
possess higher quality.
Similar conclusions can also be made by checking model textures visually.
As shown in Fig.~\ref{fig:bb_kvhd} and Fig \ref{fig:b2_kvhd}, the product
brand logos on models (Fig.~\ref{fig:bb_texture_zoom_c},
Fig.~\ref{fig:b2_texture_zoom_c}) with HD textures are clearer and sharper to
visually. However, the logos reconstructed using Kinect RGB image are quite
blurry and hard to recognise especially when the logo size is small (letters in
Fig.~\ref{fig:bb_texture_zoom_k}).

\begin{figure}[ht]
\centering
\subfigure[][]{
\includegraphics[width=.35\linewidth]{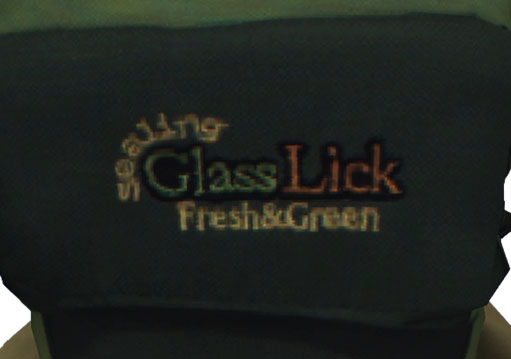}
\label{fig:bb_logo_hd_t}}
\quad
\subfigure[][]{
\includegraphics[width=.35\linewidth]{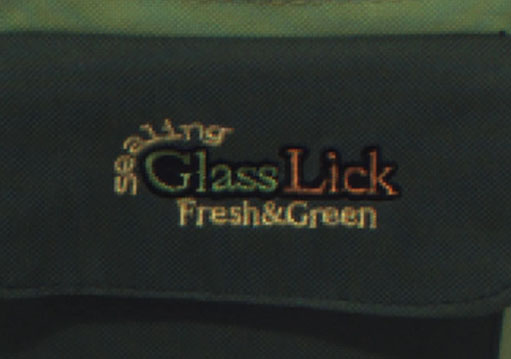}
\label{fig:bb_logo_hd_i}}
\quad
\subfigure[][]{
\includegraphics[width=.35\linewidth]{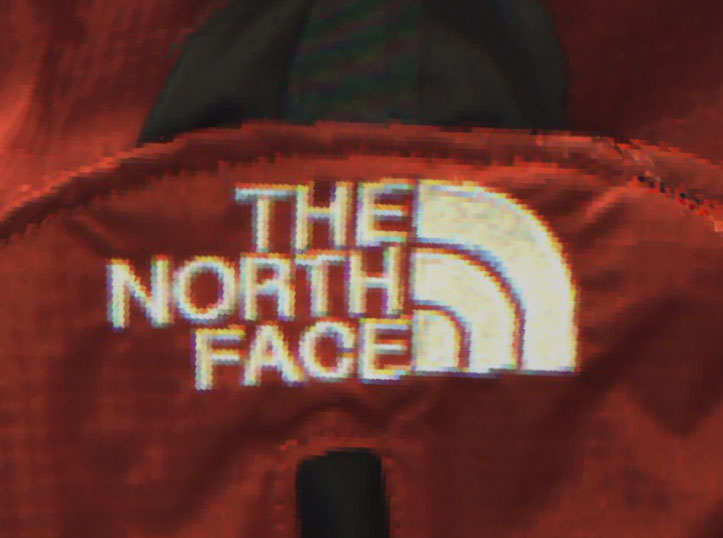}
\label{fig:b2_logo_hd_t}}
\quad
\subfigure[][]{
\includegraphics[width=.35\linewidth]{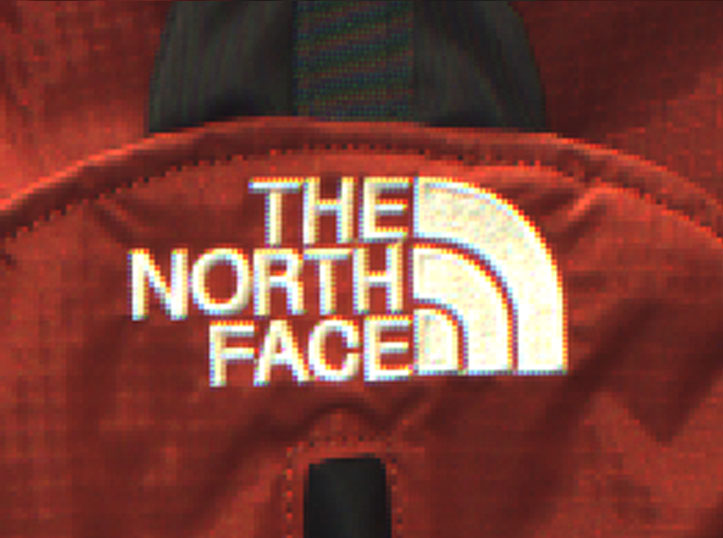}
\label{fig:b2_logo_hd_i}}
\caption{Textures using HD RGB camera vs. actual HD RGB image
(a): HD texture of a lunch bag.
(b): HD image of a lunch bag.
(a): HD texture of a backpack.
(b): HD image of a backpack.}
\label{fig:hdtvshdi}
\end{figure}

It is also worth noting that when reconstructing small-sized objects,
photorealistic textures can be achieved.
Fig \ref{fig:bb_logo_hd_t} and Fig \ref{fig:b2_logo_hd_t} are reconstructed
model textures using HD RGB image while Fig \ref{fig:bb_logo_hd_i} and Fig
\ref{fig:b2_logo_hd_i} are the corresponding parts directly cropped from HD RGB
images. Similar level of details can be observed from both model texture and
cropped HD RGB image.

\begin{table}[ht]
\small
\centering
\caption{Texture qualities (Kinect RGB vs. HD RGB)}
\label{tab:kvhd}
\begin{tabular}{|c|c|c|c|}
\hline
{\bf \begin{tabular}[c]{@{}c@{}}Image\\ Patch\end{tabular}} & {\bf \begin{tabular}[c]{@{}c@{}}Gradient\\ Magnitude\end{tabular}} & {\bf \begin{tabular}[c]{@{}c@{}}Image\\ Patch\end{tabular}} & {\bf \begin{tabular}[c]{@{}c@{}}Gradient\\ Magnitude\end{tabular}} \\ \hline
\multicolumn{4}{|c|}{{\bf Test Case: Lunch Bag}}                                                                                                                                                                                                                    \\ \hline
\multicolumn{2}{|c|}{{\bf Kinect RGB}}                                                                                      & \multicolumn{2}{c|}{{\bf HD RGB}}                                                                                    \\ \hline
Fig \ref{fig:bb_texture_zoom_k}                                                           & 0.9467                                                                  & Fig \ref{fig:bb_texture_zoom_c}                                                           & 1.8462                                                                  \\ \hline
\multicolumn{4}{|c|}{{\bf Test Case: Backpack}}                                                                                                                                                                                                                     \\ \hline
\multicolumn{2}{|c|}{{\bf Kinect RGB}}                                                                                      & \multicolumn{2}{c|}{{\bf HD RGB}}                                                                                    \\ \hline
Fig \ref{fig:b2_texture_zoom_c_k}                                                           & 4.3045                                                                  & Fig \ref{fig:b2_texture_zoom_c}                                                           & 5.8296                                                                  \\ \hline
\end{tabular}
\normalsize
\end{table}

\section{Conclusion and Discussion}
To the best of our knowledge, our method is the first aimed at creating
simplified 3D models with high quality textures.
To deliver high quality model texture, an HD RGB camera is added to work with
the depth camera of {Microsoft Kinect\texttrademark} sensor.
Improved texture update scheme on an asymmetrical colour volume with a higher
dimension than the geometry volume is presented.
Given an output decimated 3D model, our 2D texturing method is able to
maintain a high-level texture quality despite the degree of model simplification.

However, the texture quality is still limited by the dimension and size of the
colour volume which is again constrained by the GPU memory.
Our future work includes improving texture quality especially for large scene
reconstruction and exploring better 2D texture map generation methods to
achieve higher model rendering efficiency.

\section*{Acknowledgement}

This work is supported by Smart Services CRC, Australia.

{\small
\bibliographystyle{ieee}
\bibliography{3D}
}

\end{document}